\documentclass[aps,pre,floats,floatfix,twocolumn,groupedaddress,final,nobalancelastpage]{revtex4-2} 
\usepackage[T1]{fontenc}
\usepackage[utf8]{inputenc}
\usepackage{graphicx}
\usepackage{dcolumn,bm,float}
\usepackage{amssymb,amsmath,amsfonts,mathtools}
\usepackage{color,xcolor,url}
\usepackage{ragged2e}
\usepackage{booktabs}
\usepackage[percent]{overpic}
\usepackage{enumitem}
\usepackage{booktabs}
\usepackage{siunitx}
\usepackage{algorithm}
\usepackage{longtable}
\usepackage{algpseudocode}
\usepackage{xcolor} 
\usepackage{adjustbox} 
\usepackage[hidelinks]{hyperref}
\hypersetup{colorlinks=false,
    linkcolor=black,
    citecolor=black,
    urlcolor=black,
    pdftitle={Compressing regularized dynamics improves link prediction in sparse networks},
    pdfauthor={Maja Lindstr{\"o}m, Christopher Bl{\"o}cker, Tommy L{\"o}fstedt, Martin Rosvall}
}
\usepackage{orcidlink}
\usepackage{xurl}
\usepackage[capitalise]{cleveref}

\crefname{section}{Sec.}{Sections}
\crefname{Section}{Sec.}{Sections}

\setcitestyle{numbers,square,comma,sort&compress}

\makeatletter
\newcommand*{\balancecolsandclearpage}{%
  \close@column@grid
  \cleardoublepage
}
\makeatother

\usepackage{fancyhdr}
\begin{document}
\makeatletter
\renewcommand\@biblabel[1]{[#1]}
\makeatother

\title{Compressing regularized dynamics improves\\ link prediction with the map equation in sparse networks}

\author{Maja Lindstr{\"o}m\,\orcidlink{0009-0009-9224-4646}}
\email{maja.lindstrom@umu.se}
\affiliation{Department of Computing Science, MIT-huset, Ume{\aa} University, SE-901 87 Ume{\aa}, Sweden}
\affiliation{Integrated Science Lab, Ume{\aa} University, SE-901 87 Ume{\aa}, Sweden}
\affiliation{Siftlab AB, D{\"o}belnsgatan 12, SE-113 58 Stockholm, Sweden}

\author{Christopher Bl{\"o}cker\,\orcidlink{0000-0001-7881-2496}}
\affiliation{Data Analytics Group, Department of Informatics, University of Zurich, CH-8006 Zurich, Switzerland}
\affiliation{Chair of Machine Learning for Complex Networks, Center for Artificial Intelligence and Data Science (CAIDAS), University of W{\"u}rzburg, DE-97070 W{\"u}rzburg, Germany}

\author{Tommy L{\"o}fstedt\,\orcidlink{0000-0001-7119-7646}}
\affiliation{Department of Computing Science, MIT-huset, Ume{\aa} University, SE-901 87 Ume{\aa}, Sweden}

\author{Martin Rosvall\,\orcidlink{0000-0002-7181-9940}}
\affiliation{Integrated Science Lab, Ume{\aa} University, SE-901 87 Ume{\aa}, Sweden}
\affiliation{Siftlab AB, D{\"o}belnsgatan 12, SE-113 58 Stockholm, Sweden}
\affiliation{Department of Physics, Ume{\aa} University, SE-901 87 Ume{\aa}, Sweden}

\begin{abstract} 
\noindent 
Predicting future interactions or novel links in networks is an indispensable tool across diverse domains, including genetic research, online social networks, and recommendation systems. 
Among the numerous techniques developed for link prediction, those leveraging the networks' community structure have proven highly effective.
For example, the recently proposed MapSim predicts links based on a similarity measure derived from the code structure of the map equation, a community-detection objective function that operates on network flows. 
However, the standard map equation assumes complete observations and typically identifies many small modules in networks where the nodes connect through only a few links.
This aspect can degrade MapSim's performance on sparse networks.
To overcome this limitation, we propose to incorporate a global regularization method based on a Bayesian estimate of the transition rates along with three local regularization methods.
The regularized versions of the map equation compensate for incomplete observations and mitigate spurious community fragmentation in sparse networks.
The regularized methods outperform standard MapSim and several state-of-the-art embedding methods in highly sparse networks.
This performance holds across multiple real-world networks with randomly removed links, simulating incomplete observations.
Among the proposed regularization methods, the global approach provides the most reliable community detection and the highest link prediction performance across different network densities.
The principled method requires no hyperparameter tuning and runs at least an order of magnitude faster than the embedding methods.
\end{abstract}

\maketitle

\section{Introduction}
\noindent 
Predicting gene-gene interactions in biological systems paves the way for breakthroughs in genetic research~\cite{barabasi2011network}, identifying friends in online social networks enhances user engagement~\cite{song2009scalable}, and suggesting items in retail recommendation systems boosts revenue and customer satisfaction~\cite{recommender_systems}.
In these link prediction applications, both performance and interpretability are crucial.

Popular network embedding methods, such as DeepWalk~\cite{perozzi2014deepwalk}, node2vec~\cite{grover2016node2vec}, NERD~\cite{khosla}, and LINE~\cite{tang2015line}, as well as graph neural network (GNN) approaches such as GCN~\cite{kipf2016semi} and GAT~\cite{velivckovic2017graph}, predict links using a two-step approach.
First, they embed nodes in a latent space to capture the network's structure, potentially including modular patterns.
Next, they predict links based on the proximity of nodes within this latent space.
These methods are capable of capturing local and global network structures, and are also useful for node classification tasks.
While they often perform well in link prediction tests, the differences in methods and metrics for the embedding and prediction steps complicate interpretability.

The recently introduced link prediction method MapSim~\cite{pmlr-v198-blocker22a} --- map equation similarity --- offers enhanced transparency by capitalizing on the modular code structure that reflects the communities identified by Infomap~\cite{mapequation2024software}.
Infomap optimizes the flow-based and information-theoretic objective function known as the map equation to find the network partition that minimizes the modular description of a random walk on the network~\cite{smiljanić2023communitydetectionmapequation}.
The map equation favors densely connected communities where the random walk persists relatively long before leaving.
The corresponding modular code structure enables the highest compression for conveying the random walk's trajectory.
The optimal code structure best captures modular regularities in the dynamics on the network, with short codes for frequent transitions within modules and longer codes for infrequent transitions between modules.
MapSim exploits these codes by interpreting the codelength to describe a step from a given node to any other node in the network as the distance between them. 
These asymmetric node similarities enable accurate and interpretable predictions of future interactions, including novel links, in sufficiently dense networks.

Predicting links in sparse networks is more challenging because they contain less structural information.
For example, many retail recommendation systems must operate in sparse regions of networks, such as those formed by new customers or products with only a few interactions.
Since the standard map equation employed by MapSim assumes that future interactions and other unobserved links are genuinely absent, it favors many small modules to best describe the constrained dynamics in highly sparse networks, which can degrade MapSim's performance. The map equation with regularized dynamics enables more robust community detection when structural information is limited. However, whether it also leads to more robust link prediction remains unknown.

To address this gap, we employ a global regularization technique based on a Bayesian estimate of transition rates developed to identify reliable communities in networks with incomplete observations~\cite{smiljanic2021comnet}.
We also introduce three local regularization techniques: one using common-neighbor-biased transition rates and two using modified Markov times~\cite{edler2022variable}.

We found that regularizing the dynamics reduces the number of modules and helps MapSim provide more accurate link predictions based on reliable communities in highly sparse networks.
The local regularization techniques perform well on specific networks, but the global regularization performs best overall.
Basing MapSim on a Bayesian estimate of the transition rates provides a transparent, reliable, and efficient method for link prediction in sparse networks. 

\section{Background: The map equation framework}
\noindent%
The map equation~\cite{rosvall2008maps} is a flow-based objective function for community detection that combines principles from information theory and coding theory with the concept of random walks. 
The map equation measures the quality of a network partition as the expected per-step description length --- the codelength --- for a random walk on the network.
Detecting the optimal communities is a search problem that involves minimizing the map equation by identifying sets of nodes where the random walker stays for a relatively long time.

\begin{figure}[tp!]
    \centering
    \begin{overpic}[width=\linewidth]{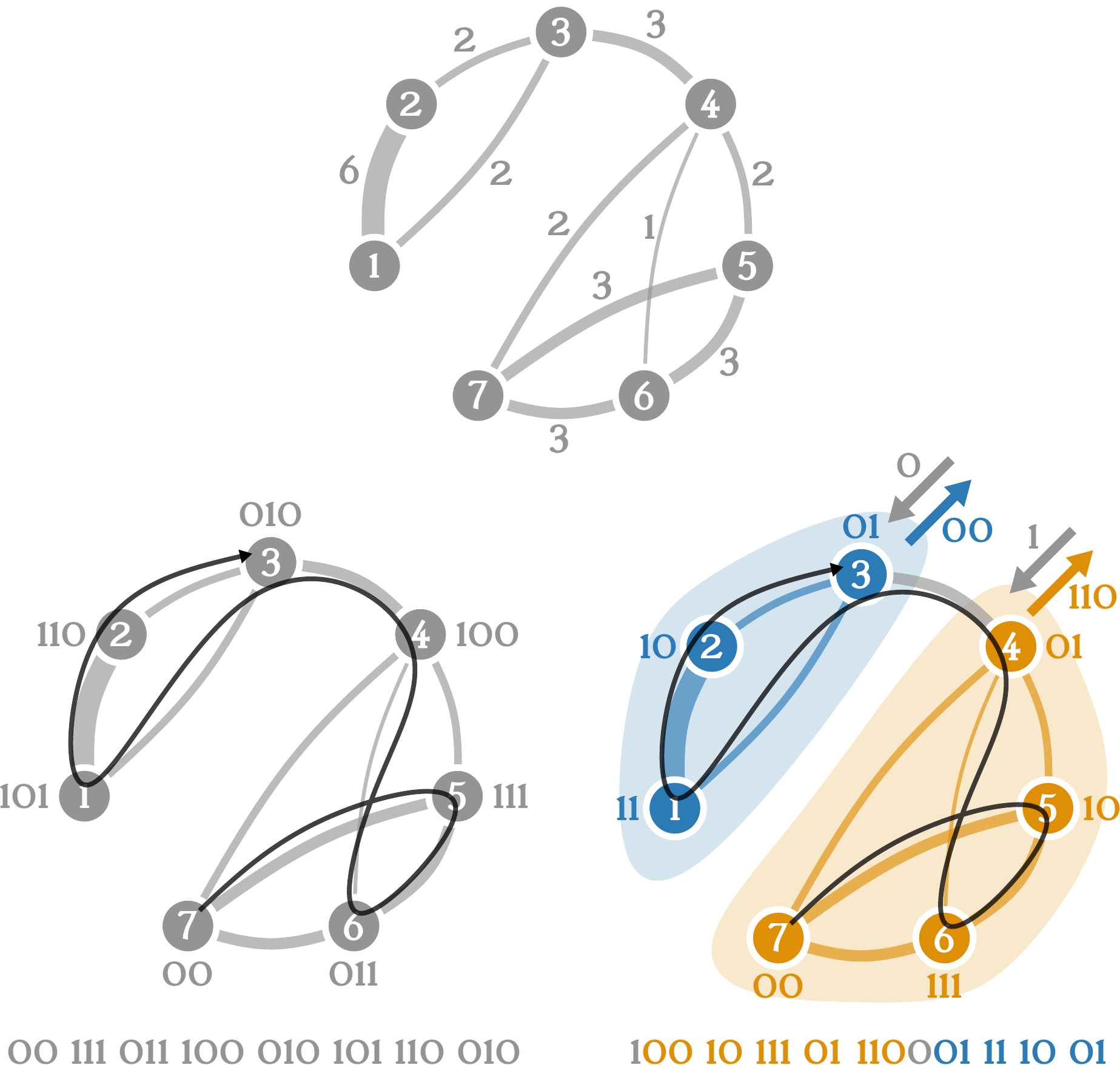}
        \put(27,92){\textbf{(a)}}
        \put(0,45){\textbf{(b)}}
        \put(54,45){\textbf{(c)}}
    \end{overpic}
    \caption{\textbf{A communication game on a network.} The black line illustrates a possible path of a random walk on the network, with colors representing different modules. \textbf{(a)} An example network with 7 nodes and 10 links. Link widths correspond to their weights, shown next to each link. \textbf{(b)} A one-level partition with all nodes grouped into the same module, and seven unique codewords. The sequence of codewords below the network describes the random walk ($23$ bits). \textbf{(c)} A two-level partition where nodes are divided into two modules with reusable codewords. Arrows show codewords for entering and exiting modules. The random walker's path can now be communicated more efficiently ($22$ bits).}
    \vspace{-1em}
    \label{fig:map-equation-principle}
\end{figure}

For illustration, consider a communication game where the sender observes a random walker on a network and updates the receiver about the random walker's position after each step.
The network $G = \left(V, E\right)$ consists of nodes $V$ and possibly weighted and directed links $E$, with $w_{uv}$ denoting the weight on the link from node $u$ to node $v$ (\cref{fig:map-equation-principle}a).
Given an assignment of nodes to modules, how succinctly can the sender update the receiver about the position of the random walker?
Which assignment gives the shortest description?

In the simplest case, all nodes belong to the same community and receive unique codewords, for example using Huffman coding~\cite{huffman1952method}, where more frequently visited nodes receive shorter codewords (\cref{fig:map-equation-principle}b).
According to Shannon's source coding theorem~\cite{shannon1948mathematical}, the shortest possible codelength is the entropy of the nodes' visit rates
\begin{equation}
    L = \mathcal{H}(P)= - \sum_{u\in V} p_u \log_2 p_u,
    \label{codelength_M1}
\end{equation}
where $\mathcal{H}$ is the Shannon entropy, $P = \left\{p_u \mid u \in V \right\}$ is the set of the nodes' visit rates, and $p_u$ is the visit rate of node $u$.

In networks with community structure, the description of the random walker's path can be compressed by partitioning the nodes into modules $\mathsf{m} \in \mathsf{M}$. 
Within each module $\mathsf{m}$, nodes receive unique codewords, but the same codewords can be reused in different modules, leading to shorter codewords on average (\cref{fig:map-equation-principle}c).
To communicate when the random walker switches modules, each module also receives a dedicated exit codeword, and entering modules is described using a designated index-level codebook.
The two-level map equation calculates the codelength for a partition of the nodes into modules $\mathsf{M}$ as an average of index-level and module-level entropies, weighted by the rates at which the respective codebooks are used
\begin{equation}
    L(\mathsf{M}) = q \mathcal{H}(Q) + \sum_{\mathsf{m} \in \mathsf{M}}p_\mathsf{m} \mathcal{H}(P_\mathsf{m}),
    \label{map_equation}
\end{equation}
where $q = \sum_{\mathsf{m} \in \mathsf{M}}q_\mathsf{m}$ is the index codebook's use rate and $q_\mathsf{m} = \sum_{u \notin \mathsf{m}} \sum_{v \in \mathsf{m}} p_u t_{uv}$ is the entry rate for module $\mathsf{m} \in \mathsf{M}$ with $t_{uv} = {w_{uv}}/{\sum_{v' \in V} w_{uv'}}$ for the transition probability from $u$ to $v$.
The set of normalized module entry rates is $Q = \{{q_\mathsf{m}}/{q} \mid \mathsf{m} \in \mathsf{M}\}$, and the use rate for module $\mathsf{m}$'s codebook, $p_\mathsf{m}$, is the sum of the exit rate for $\mathsf{m}$ and the visit rates of all nodes in $\mathsf{m}$, $p_\mathsf{m} = \mathsf{m}_\text{exit} + \sum_{u \in \mathsf{m}}p_u$, where $\mathsf{m}_\text{exit} = \sum_{u \in \mathsf{m}} \sum_{v \notin \mathsf{m}} p_u t_{uv}$.
Finally, $P_\mathsf{m} = \{{\mathsf{m}_\text{exit}}/{p_\mathsf{m}}\} \cup \{{p_u}/{p_\mathsf{m}} \mid u \in \mathsf{m} \}$ is the set of module-normalized node visit rates in $\mathsf{m}$, including $\mathsf{m}$'s exit rate.
Through recursion, the map equation has been generalized for hierarchical partitions with nested submodules~\cite{multilevel_mapeq}.

\subsection{Global regularization}
\begin{figure}[tp!]
    \centering
    \begin{overpic}[width=\linewidth]{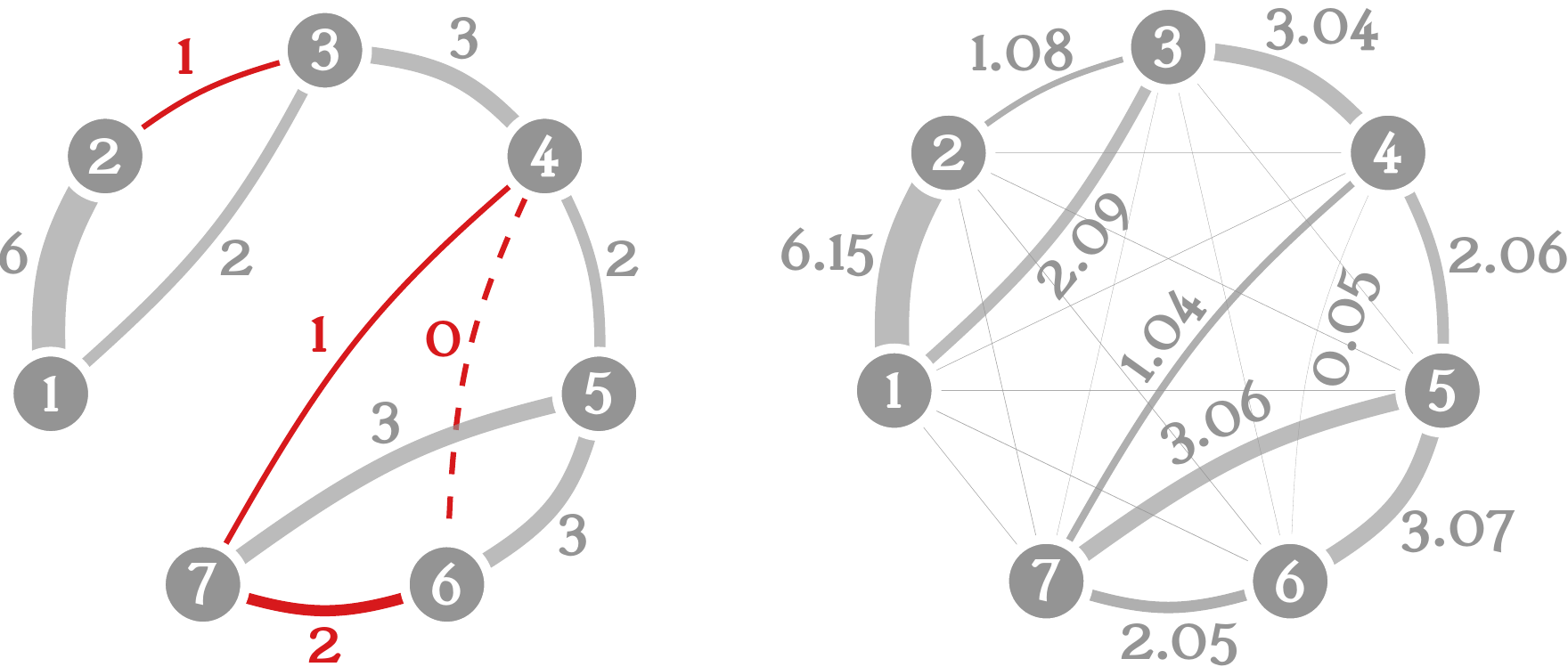}
        \put(0,40){\textbf{(a)}}
        \put(53,40){\textbf{(b)}}
    \end{overpic}
    \caption{\textbf{Global regularization.} \textbf{(a)} A weighted network with incomplete observations, where missing links are highlighted in red. \textbf{(b)} The network after applying global regularization, incorporating a Bayesian estimate of the transition rates to adjust the flows.}
    \vspace{-1em}
    \label{fig:global-regularization-example}
\end{figure}

\noindent%
The standard map equation assumes complete observations and considers unobserved links as genuinely absent, which can cause issues in networks with missing observations.
For reliable community detection in such cases, the map equation can be regularized using a Bayesian estimate of the transition rates between nodes~\cite{smiljanic2021comnet}.
This approach considers networks with multi-edges where the observed integer-valued link counts can be smaller or equal to the true counts (\cref{fig:global-regularization-example}).

The transition rate between nodes $u$ and $v$, denoted $t_{uv}$, is estimated by introducing a prior over the probabilities that a random walker steps from node $u$ to other nodes, $T_u = (t_{u1},\ldots,t_{un})$, where $n = \lvert V \rvert$ is the number of nodes in the network.
The posterior transition rates are estimated as
\begin{equation}
    \hat t_{uv}(W_u) = \int t_{uv} P(T_u \mid W_u) \,dT_u,
    \label{eq:posterior_transition_rates}
\end{equation}
where $W_u = (w_{u1},\ldots, w_{un})$ and $w_{uv}$ is the number of observed links between nodes $u$ and $v$.
The posterior $P\left(T_u \mid W_u\right)$ is determined using a Dirichlet prior, and the likelihood
\begin{equation}
    P\left(W_u \mid T_u\right) = s_u^{\mathrm{out}}! \prod_{v=1}^{n} \frac{t_{uv}^{w_{uv}}}{w_{uv}!},
    \label{likelihood}
\end{equation}
where $s_u^{\mathrm{out}} = \sum_{v = 1}^{n} w_{uv}$ is the out-strength of node $u$.
The estimated transition rates in \cref{eq:posterior_transition_rates} can be computed as
\begin{align}
    \hat t_{uv}(W_u) & = \frac{w_{uv} + \gamma_{uv}}{\sum_{v=1}^{n} w_{uv} + \gamma_{uv}} \nonumber \\
    & = (1-\alpha_u) \frac{w_{uv}}{\sum_v w_{uv}} + \alpha_u \frac{\gamma_{uv}}{\sum_v \gamma_{uv}},
    \label{eq:transition_rates}
\end{align}
where $\alpha_u = \frac{\sum_{v=1}^{n} \gamma_{uv}}{\sum_{v=1}^{n} w_{uv} + \gamma_{uv}}$ can be interpreted as a node-dependent teleportation parameter for the random walker.
A random walker at node $u$ follows an observed link with probability $\left(1-\alpha_u\right)$, corresponding to the first term in \cref{eq:transition_rates}, and a link in the prior network with probability $\alpha_u$, corresponding to the second term.
The parameters $\gamma_{u1},\ldots, \gamma_{un}$ are link weights in the prior network defined as
\begin{equation}
    \gamma_{uv} = \frac{\ln (n)}{n} c_{uv},
    \label{eq:gamma}
\end{equation}
where $c_{uv}$ is the expected link weight between nodes $u$ and $v$ according to the so-called continuous configuration model~\cite{palowitch2018significance}. It is defined as
\begin{equation}
    c_{uv} = \frac{\sum_{i \in V} k_i^{\text{in}} + k_i^{\text{out}}}{\sum_{i \in V} s_i^{\text{in}} + s_i^{\text{out}}} \frac{s_u^{\text{out}} s_v^{\text{in}}}{k_u^{\text{out}} k_v^{\text{in}}},
    \label{link_weight}
\end{equation}
where $k_u^{\text{in}}$ and $k_u^{\text{out}}$ denote observed in- and out-degrees, and $s_u^{\text{in}} = \sum_{v = 1}^{n} w_{vu}$ is the in-strength for node $u$.

Infomap can generate the weighted network with global regularization, $\textbf{W}_{\operatorname{g-reg}}$, at no extra computational cost thanks to an efficient representation of the prior network.

\subsection{MapSim}
\begin{figure}[tp!]
    \centering
    \begin{overpic}[width=\linewidth]{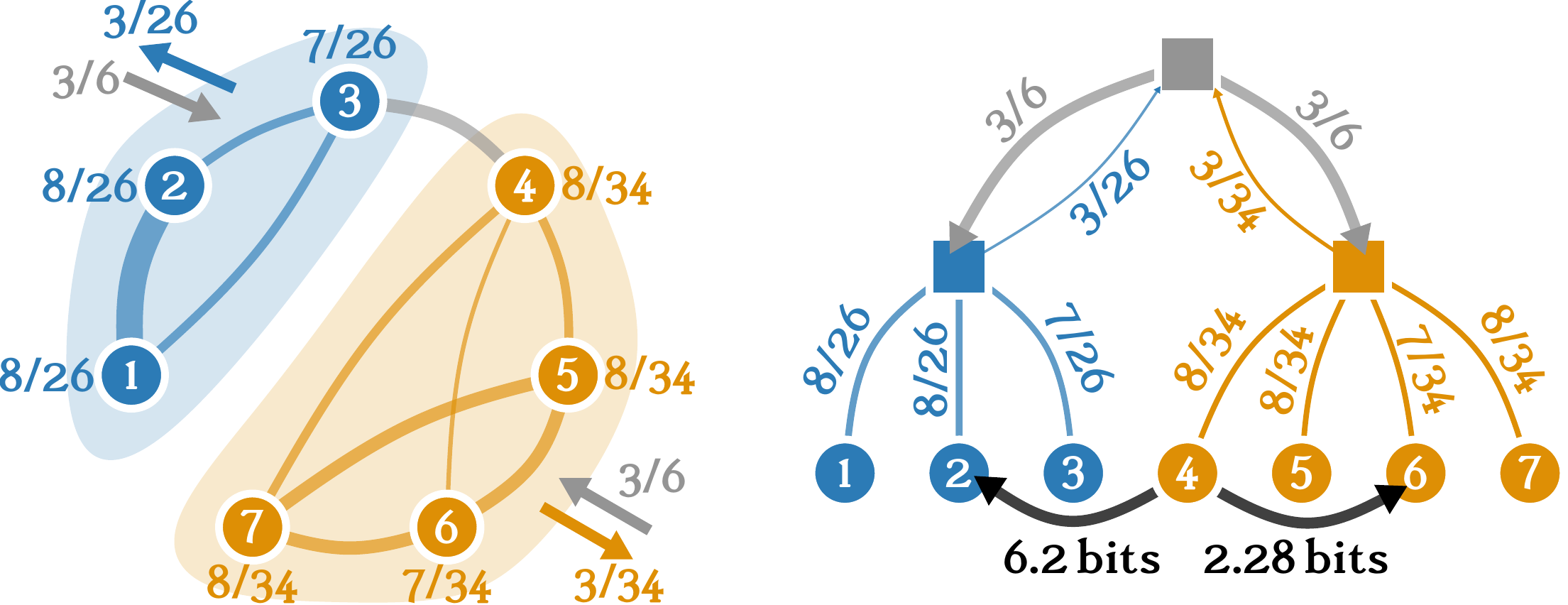}
        \put(0,40){\textbf{(a)}}
        \put(53,40){\textbf{(b)}}
    \end{overpic}
    \caption{\textbf{MapSim.} \textbf{(a)} A network with two modules and transition rate annotations for moving in the network. \textbf{(b)} A tree-like representation of the network, annotated with transition rates, showing the cost in bits for moving from node 4 to 2, and from node 4 to 6, respectively.}
    \vspace{-1em}
    \label{fig:mapsim-principle}
\end{figure}

\noindent%
MapSim is a node-similarity measure based on the map equation introduced for similarity-based link prediction~\cite{pmlr-v198-blocker22a}.
MapSim uses the map equation's code structure to compute pairwise similarity scores between nodes (\cref{fig:mapsim-principle}).
Nodes in the same module --- or submodule in hierarchical partitions --- are generally considered more similar,  as transition between them can be described more efficiently using shorter codewords, corresponding to more probable links.
To compute the similarity between nodes $u$ and $v$, MapSim identifies the smallest module $\mathsf{m} \in \mathsf{M}$ that contains both nodes. In a hierarchical partition, this means the lowest-level submodule that includes both $u$ and $v$, whereas in a two-level partition, it simply corresponds to the shared module of the nodes.
With rate $\operatorname{rev}(\mathsf{m}, u)$ at which a random walker at $u$ travels to $\mathsf{m}$'s root, and rate $\operatorname{fwd}(\mathsf{m}, v)$ at which a random walker at the root of $\mathsf{m}$ travels to $v$,
the cost in bits for stepping from $u$ to $v$ is
\begin{equation}
    \text{MapSim}(\mathsf{M}, u, v) = -\log_2(\operatorname{rev}(\mathsf{m}, u) \cdot \operatorname{fwd}(\mathsf{m}, v)).
    \label{eq:mapsim}
\end{equation}
With this approach, MapSim can describe any link, whether that link exists in the network or not.
This capability is crucial in settings such as recommendation systems, where many links have yet to form.
By relating the probability that nonlinks form to their description cost, MapSim predicts future links.

\section{Regularization for link prediction}
\noindent%
Link observations are incomplete in many real-world scenarios for various reasons:
some links are harder to observe, others have not yet formed, or the network is too costly to observe in its entirety.
Applying the standard map equation to networks with incomplete observations can lead to inaccurate communities if the observed link weights reflect the true weights poorly, affecting downstream tasks such as link prediction with MapSim.
To account for incomplete observations, we propose to combine MapSim with the global regularization of the random walker's transition rates.

\begin{figure*}[htp!]
    \centering
    \begin{overpic}[width=\linewidth]{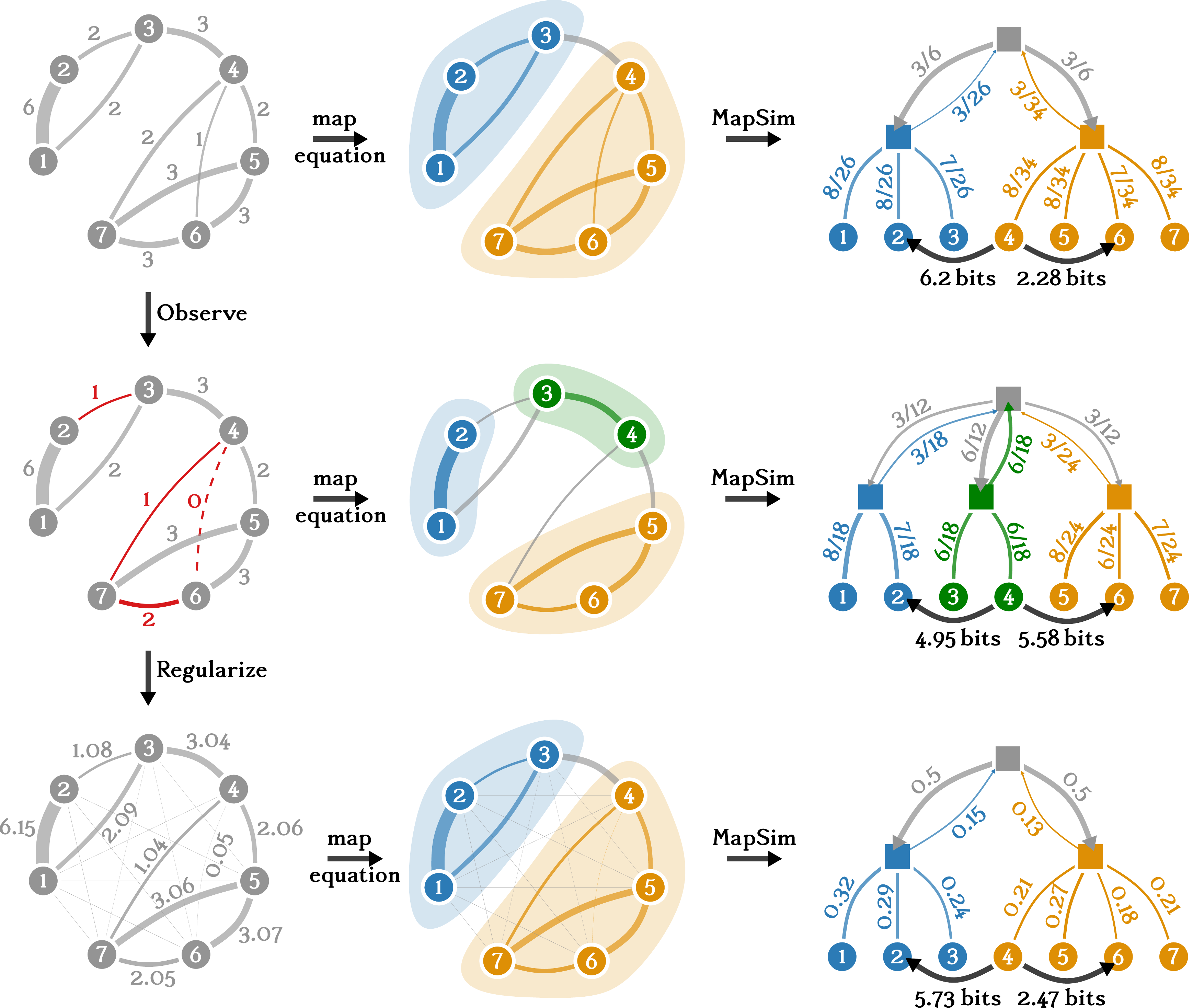}
        \put(0,82){\textbf{(a)}}
        \put(0,52){\textbf{(b)}}
        \put(0,22){\textbf{(c)}}
    \end{overpic}
    \caption{\textbf{Combining MapSim with a Bayesian estimate of the transition rates for reliable link prediction in incomplete networks.}
    \textbf{(a)} The complete network with line widths representing link weights. The map equation identifies two communities. With MapSim, the network is represented as a hierarchical coding tree with transition rates shown between the nodes. The bit costs of transitioning from node 4 to nodes 2 and 6 are shown, with the most probable link between nodes 4 and 6.
    \textbf{(b)} The observed network where some link weights are observed incorrectly, indicated in red. The map equation now identifies three communities, and with MapSim, the link between nodes 4 and 2 becomes more likely.
    \textbf{(c)} The regularized map equation correctly detects the original two communities, restoring MapSim's prediction that the link between nodes 4 and 6 is the most likely. 
    }
    \vspace{-1em}
    \label{fig:global_reg}
\end{figure*}

The schematic example in \cref{fig:global_reg} illustrates how incomplete observations distort link predictions and how global regularization mitigates this effect.
With complete data, the standard map equation detects two communities (\cref{fig:global_reg}a).
After removing some of the multi-edges to simulate incomplete data, it identifies three communities, altering the link prediction results (\cref{fig:global_reg}b).
In the complete network, MapSim considers a link from node 4 to~6 more likely than one from node 4 to 2 due to the lower description cost.
Conversely, with incomplete data, the link between nodes 4 and 2 becomes more likely.
Applying global regularization recovers the two communities, restoring MapSim's prediction that the link from node 4 to 6 is more likely (\cref{fig:global_reg}c).

In some cases, global regularization can obscure local regularities in networks.
In response, we evaluated three local regularization techniques: two based on established methods and one using a new approach that reconnects nodes based on their proximity to restore local regularities in incomplete networks.
Common Neighbors focuses on connecting nodes that share neighbors.
Mixed Markov Time (MMT) works similarly but extends naturally to weighted links.
The proposed new method, Variable Markov Time (VMT), adjusts the random walker's speed according to node degrees, allowing it to move farther in sparser regions~\cite{edler2022variable}.
By adding links preferentially in these regions, it compactifies chains of low-degree nodes that can cause spurious modules while leaving denser areas intact.
The local regularization methods generate new weighted links.
We combined the new weights from the local regularization, $\textbf{W}_{\operatorname{l-reg}}$, with those from the global regularization, $\textbf{W}_{\operatorname{g-reg}}$, by summing their weights
\begin{equation}
    \textbf{W}_{\operatorname{reg}}=\textbf{W}_{\operatorname{g-reg}}+\textbf{W}_{\operatorname{l-reg}}.
    \label{eq:weight_sum}
\end{equation}

\subsection{Local Regularization --- Common Neighbors}
\begin{figure}[htp!]
    \centering
    \begin{overpic}[width=\linewidth]{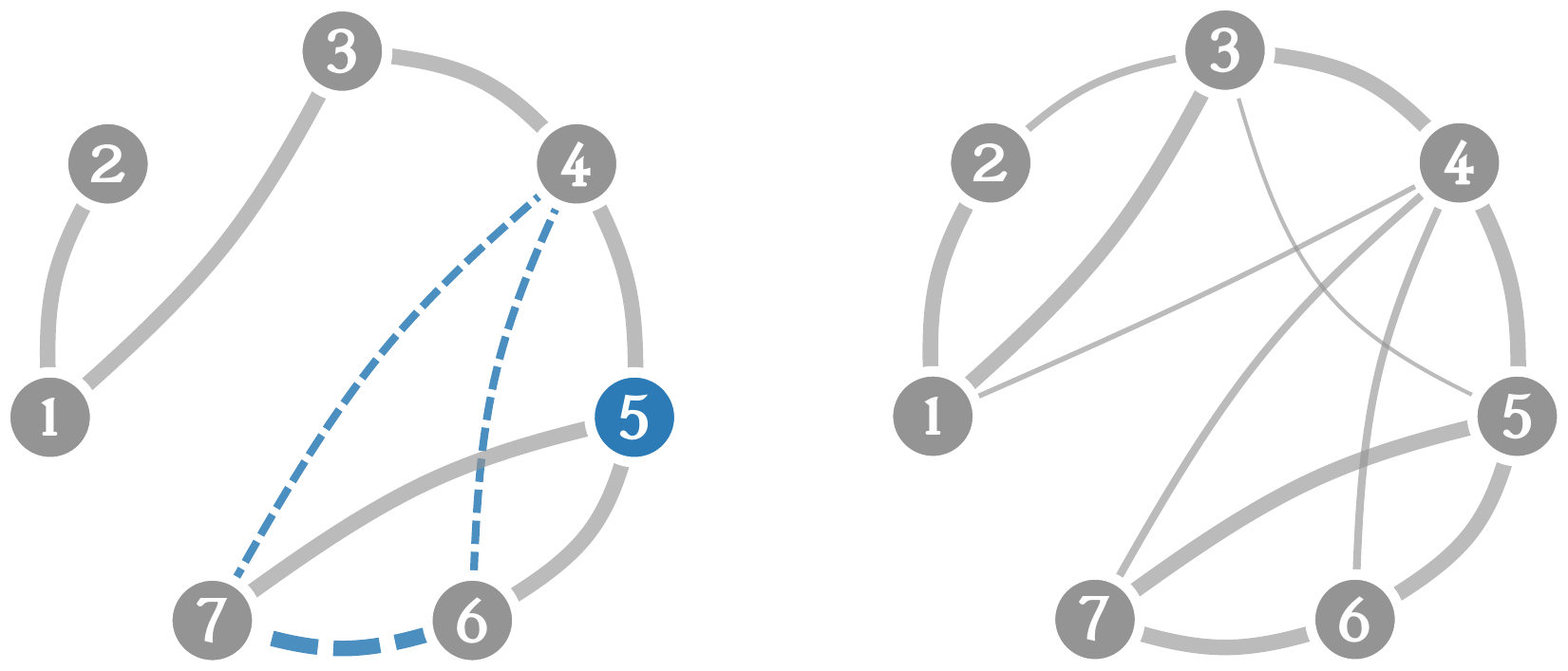}
        \put(0,40){\textbf{(a)}}
        \put(53,40){\textbf{(b)}}
    \end{overpic}
  \caption{\textbf{Common Neighbors.} \textbf{(a)} The unweighted example network with missing observations. This example is based on node 5, highlighted in blue. Node 5 is a common neighbor of nodes 4 and 6, nodes 4 and 7, and nodes 6 and 7. If two nodes have a common neighbor, they are connected by a link, shown with dashed blue lines. \textbf{(b)} The resulting network. For simplicity, link widths represent the average weight of the two directed links between each node pair.}
  \label{fig:schematic_common_neighbours}
\end{figure}

\noindent%
Common Neighbors connects closely related nodes in triangles~\cite{lu2011link}.
If two nodes, $u$ and $v$, have a common neighbor, they are connected by a link with weight
\begin{equation}
     w_{\text{uv}} = \frac{\lvert N(u) \cap N(v) \rvert}{\lvert N(u) \cup N(v) \rvert},
    \label{eq:common_neighbours}
\end{equation}
where $N(u)$ and $N(v)$ are the sets of adjacent nodes to $u$ and $v$, respectively.
Common Neighbors then constructs a matrix $\textbf{W}_{\operatorname{l-reg}}$ where the entries are defined in \cref{eq:common_neighbours}.
Applied to the example network with missing observations in \cref{fig:global-regularization-example} with unweighted links for simplicity, Common Neighbors reconstructs some of the missing links (\cref{fig:schematic_common_neighbours}).

\subsection{Local Regularization --- Mixed Markov Time}
\begin{figure}[htp!]
    \centering
    \begin{overpic}[width=\linewidth]{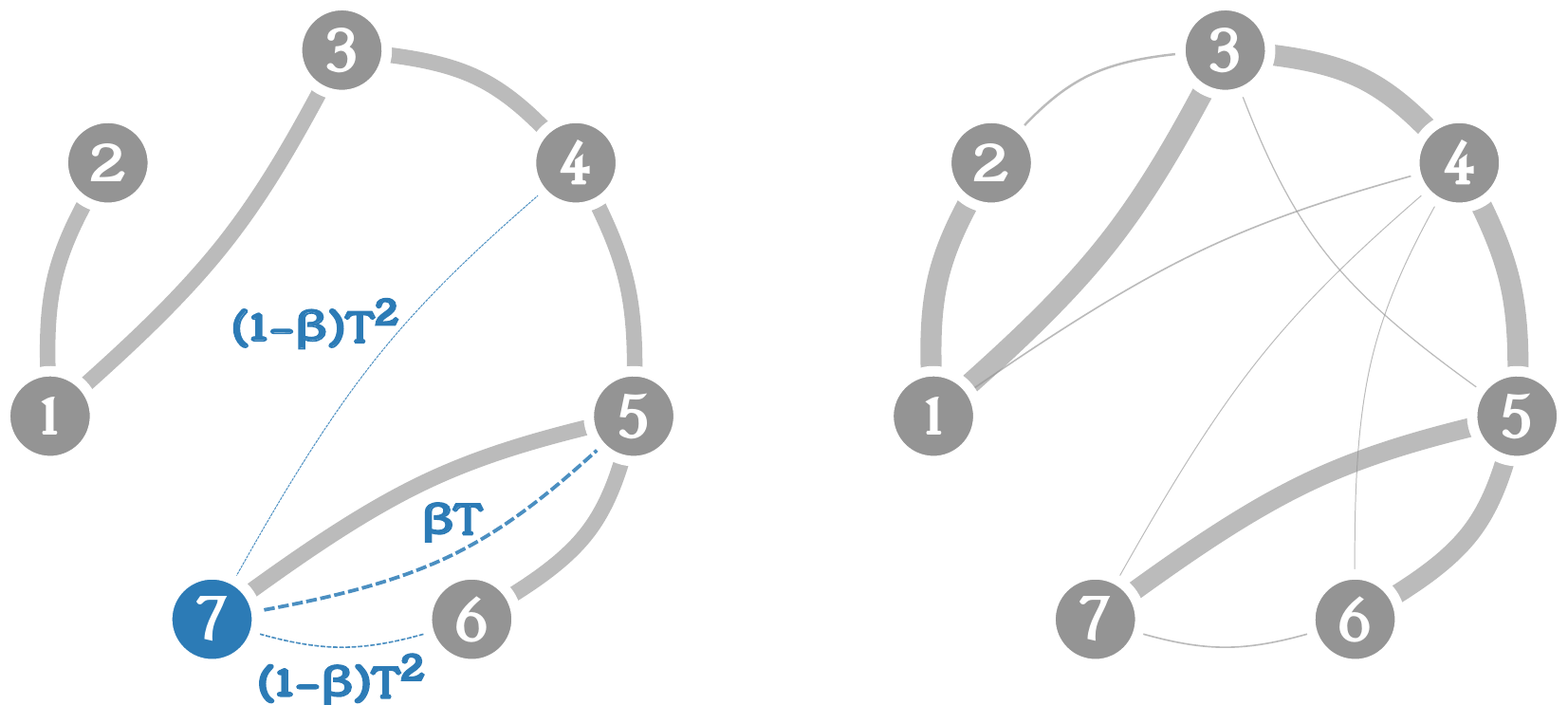}
        \put(0,40){\textbf{(a)}}
        \put(53,40){\textbf{(b)}}
    \end{overpic}
  \caption{\textbf{Mixed Markov Time.} The unweighted example network with missing observations. We base this example on node 7, highlighted in blue. The added links from node 7 are shown as dashed blue lines. These contributions arise either from its immediate neighbors via the transition matrix, $\textbf{T}$, or from nodes two steps away through the squared transition matrix, $\textbf{T}^2$, as indicated on the links. Here we use $\beta=0.7$. \textbf{(b)} The resulting network. For simplicity, link widths represent the average weight of the two directed links between each node pair.}
  \label{fig:schematic_MMT}
\end{figure}

\noindent%
Missing links can create chains of low-degree nodes that prevent random walkers from reaching otherwise nearby nodes.
We address this effect by taking longer steps in the network using the squared transition matrix, $\textbf{T}^2$, which connects second-degree neighbors.
To form the reconstructed network, we combine the existing transitions, represented by the transition matrix $\textbf{T}$, and the two-step transitions, represented by the squared transition matrix $\textbf{T}^2$, weighted by the parameter $\beta$
\begin{equation}
    \textbf{W}_{\operatorname{l-reg}} = \beta \textbf{T} + (1-\beta)\textbf{T}^2,
    \label{eq:MMT}
\end{equation}
where we set the matrix diagonals to zero to avoid self-links, as they would counteract the effect of enabling longer steps.
\Cref{fig:schematic_MMT} illustrates the Mixed Markov Time regularization with the incomplete example network from \cref{fig:global-regularization-example} with unweighted links for simplicity.
Like Common Neighbors, this regularization method also reconstructs some of the missing links.

\subsection{Local Regularization --- Variable Markov Time}

\noindent%
The map equation formulates community detection as a compression problem by modeling network flow as a random walk.
Minimizing the random walk's description corresponds to identifying communities that characterize the network's organization.
In regions made sparse by missing link observations, such as chains of nodes, this approach may lead to over-partitioning.
To reduce this effect, we propose a local regularization approach based on Variable Markov Time scaling~\cite{edler2022variable}.
By adding links between nearby low-degree nodes, a random walker will move faster in sparse regions and slower in dense ones, reducing over-partitioning.

\begin{figure}[htp!]
    \centering
    \begin{overpic}[width=\linewidth]{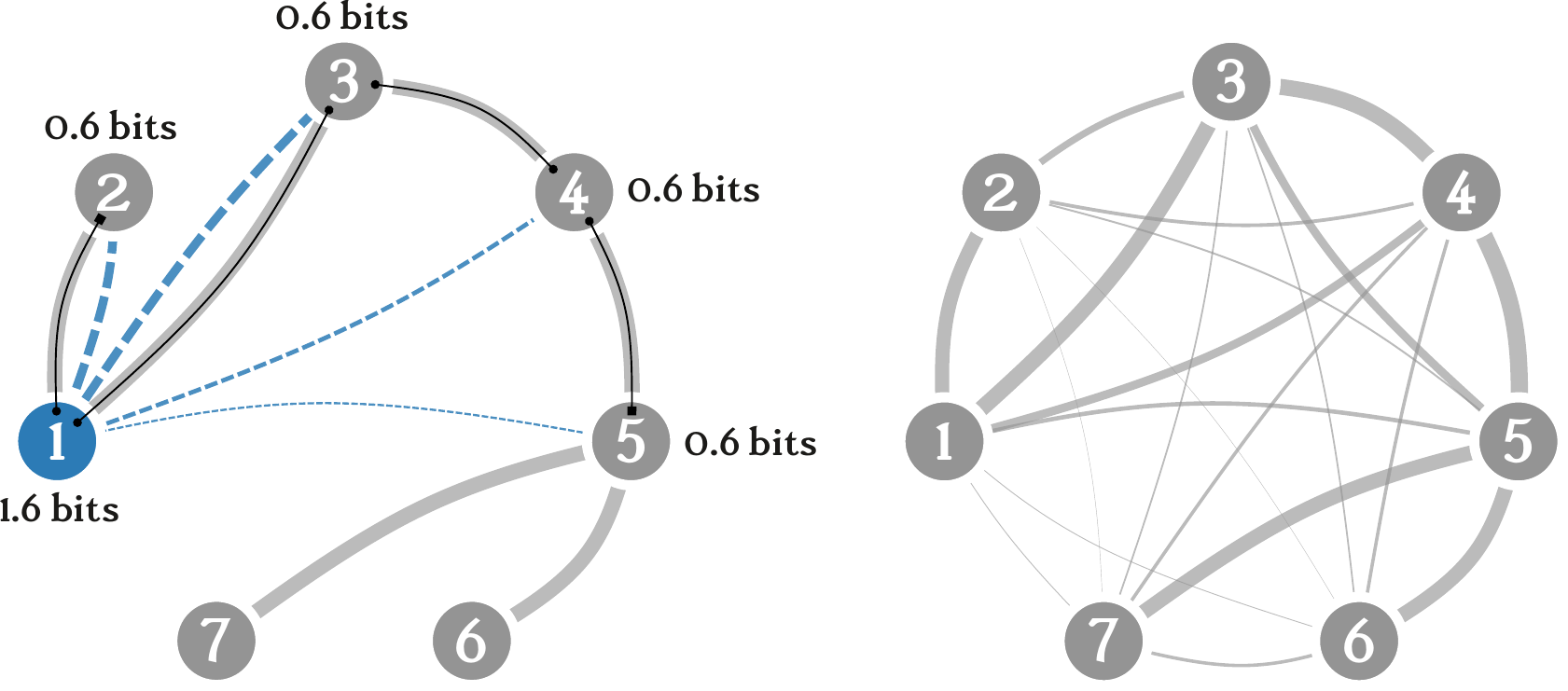}
        \put(0,40){\textbf{(a)}}
        \put(53,40){\textbf{(b)}}
    \end{overpic}
    \caption{\textbf{Variable Markov Time.} \textbf{(a)}
    The unweighted example network with missing observations. The blue node 1 indicates an example where the random walker starts its walk. The black lines show the paths of the random walker and the dashed blue lines illustrate the new links. \textbf{(b)} The resulting network. For simplicity, link widths represent the average weight of the two directed links between each node pair.
    }
  \label{fig:schematic_variable_markov}
\end{figure}

To implement Variable Markov Time scaling, we preprocess the network and modify the random walker's dynamics.
We begin by assigning the random walker a budget $b = \log_2(k_\text{max})$, where $k_\text{max}$ is the maximum node degree in the network.
This budget represents the walker's capacity to explore the network.
To reconstruct the network locally, we consider each node $u \in V$ individually.
The recursive nature of this method allows the random walker to explore multiple paths starting from node $u$.
For example, consider the unweighted example network from \cref{fig:global-regularization-example} with node 1 as the starting point for the random walker (\cref{fig:schematic_variable_markov}). The highest node degree is 3, giving the walker a budget of $\log_2(3) \approx 1.6$ bits. First, the walker evaluates the cost of taking a step, which for unweighted links is based on the number of outgoing links $k$ from node $u$, requiring $\log_2 k$ bits. When dealing with weighted links, the cost is calculated using the Shannon entropy, $\text{cost} = \mathcal{H}\left(\left\{w_{uv} / s_u^\text{out} \mid v \in N\left(u\right)\right\}\right)$. In this example, the walker starts at node 1 and takes a step to node~3. The cost of this step is thus $\log_2 2 = 1$ bit, reducing the remaining budget accordingly. A new link weight between node~1 and node 3 is added.
At each step, the random walker decides with an absorption probability $\delta$ whether to stop or continue. If the walker decides to continue, it checks if the remaining budget is sufficient to take another step. If so, it moves to a neighboring node, $v$; in our example, node 4. The cost of taking this step is 0 bits because the walker does not have the option to return to previously visited nodes along its current path. A link between node~1 and node 4 is added, and the process repeats until either the walker exhausts its budget or stops due to the absorption probability.
Once the walker has completed one path, it resets and explores another. For example, in the next iteration, the walker takes a step from node 1 to node 2, which costs 1 bit. This process continues for all possible paths from node $u$, generating flows.
We multiply these new flows with $u$'s flow, $p_u$, and the network's total weight $w_\text{tot} = \sum_{u, v \in V} w_{uv}$.
In this way, the resulting weights of this local regularization method, $\textbf{W}_{\operatorname{l-reg}}$, have the same total weight as in the observed network.

\section{Results and Discussion}
\noindent%
To evaluate the link prediction methods on sparse networks, we simulated incomplete data by randomly removing different fractions of links, $r \in \left\{0.1, 0.2, \ldots, 0.9\right\}$. The remaining links form subsets of the original networks, which we refer to as training networks.
The sets of removed $x$ links form the validation sets, representing the true links the models aimed to recover.
For each validation set, we also sampled $x$ nonlinks uniformly and at random as negative examples to ensure balanced comparisons between existing and nonexistent links during prediction.
For undirected networks, we considered links in both directions and sampled twice as many nonlinks.
To improve statistical reliability, we generated five training networks for each fraction of removed links.
Throughout the evaluation process, we ensured that all nodes were retained, even those left without links.

We explored different ways to apply regularization, including one global and three local regularization techniques: Common Neighbors, Mixed Markov Time, and Variable Markov Time.
To ensure consistent regularization across networks of different sizes, we introduced a finite-size correction, replacing $n$ with $n+C$ in the original Bayesian prior strength $\ln(n)/n$ in \cref{eq:gamma} derived for large networks \cite{smiljanic2021comnet}. The adjusted prior strength, $\ln(n+C)/(n+C)$ with $C=50$, preserves meaningful community structures in small networks while maintaining the correct large-$n$ behavior. Since different regularization methods affect the network density, we report the resulting total link weight for each fraction of removed links and regularization method in \cref{tab:density}. 

We analyzed the link prediction performances on 38 real-world networks from various fields~\cite{peixoto2020netzschleuder}, including directed, weighted, undirected, and unweighted examples (\cref{tab:dataset_summary}).
We evaluated the performance by the mean Area Under the Receiver Operating Characteristic Curve (AUC) for each fraction of removed links. The datasets are balanced with the same number of positive and negative links, making AUC more suitable than the Area Under the Precision-Recall Curve (AUPRC)~\cite{saito2015precision}. 
We compared the AUC scores for the different versions of MapSim with the following state-of-the-art link prediction methods: node2vec~\cite{grover2016node2vec}, NERD~\cite{khosla}, LINE$_1$, LINE$_2$, LINE$_{1+2}$~\cite{tang2015line}, GCN~\cite{kipf2016semi}, and GAT~\cite{velivckovic2017graph}.

Due to the variability in how various link prediction methods perform across different networks, we performed hyperparameter tuning using Hyperopt~\cite{bergstra2013making}, a Bayesian optimization framework that uses the Tree-structured Parzen Estimator (TPE) to efficiently search the hyperparameter space. We used the hyperparameter spaces recommended by the authors, setting a maximum of 20 evaluations per method, with the objective to minimize the AUC score.
For node2vec, we conducted $r \in \left\{60, 70, 80, 90, 100\right\}$ random walks of length $l \in \left\{30, 40, 50, 60\right\}$ per node and employed a window size of $w \in \left\{5, 10, 15\right\}$.
We set the in-out parameter $p$, and the return parameter $q$, as $p,q \in \left\{0.5, 1, 2, 4\right\}$.
For LINE, we executed first-order (LINE$_1$), second-order (LINE$_2$), and combined first-and-second-order proximity (LINE$_{1+2}$). We created $s \in \left\{2, 3, 4, 5\right\}$ negative samples and set the starting value of the learning rate to $\rho \in \left\{0.0001, 0.001, 0.01\right\}$.
For NERD, we used the same hyperparameter space as for LINE, with the additional parameter walk size set to $n \in \left\{1, 2, 3, 4, 5\right\}$.
For the graph neural network methods, GCN and GAT, we set the hidden dimensions to $h \in \left\{32, 64, 128\right\}$, the number of layers to $L \in \left\{1, 2, 3\right\}$, the out dimension to $d_{out} \in \left\{64, 128, 256\right\}$ and finally, the learning rate to $\rho \in \left\{0.00001, 0.0001, 0.001, 0.01\right\}$.
All embeddings were configured with a dimension size of $d \in \left\{64, 128, 256\right\}$. The optimal hyperparameters for each method and fraction of removed links used in our experiments are detailed in the Appendix (\cref{tab:hyperparams_Node2Vec}, \cref{tab:hyperparams_NERD}, \cref{tab:hyperparams_LINE1}, \cref{tab:hyperparams_LINE2}, \cref{tab:hyperparams_LINE12}, \cref{tab:hyperparams_GCN}, and \cref{tab:hyperparams_GAT}).

For MapSim, we used the map equation optimization algorithm Infomap~\cite{mapequation2024software} to detect the optimal communities where no hyperparameters needed to be tuned.
However, we ran it 100 times to avoid local minima and picked the partition with the shortest codelength.
For the embedding-based approaches, we computed similarity scores by taking the sigmoid of the dot product of the nodes' embedding vectors, as proposed by Khosla et al.~\cite{khosla}.
For MMT, we set $\beta$ to $0.7$ based on link prediction tests for different values of $\beta$.
This value corresponds to the random walker taking $1.3$ steps on average, balancing between existing links and allowing random walks to escape isolated areas.
Varying $\beta$ around this value had little effect on the average AUC scores~(\cref{tab:beta_MMT}).
For VMT, larger $\delta$ values favored stronger short-range links, while smaller values strengthen long-range links.
We set $\delta = 0.5$ to balance short- and long-range links with minimal assumptions.

\begin{figure*}[htp!]
    \centering
    \includegraphics[width=0.97\textwidth]{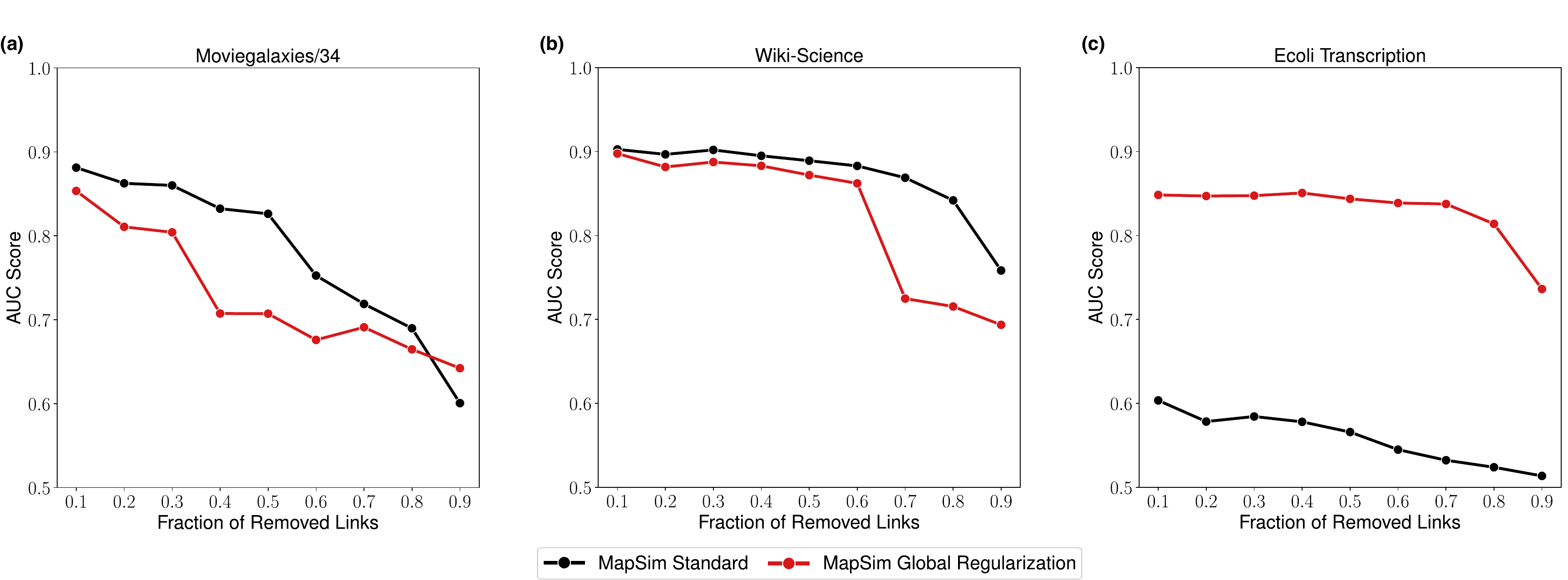}
    \caption{\textbf{Link prediction performance of standard MapSim and MapSim with global regularization.} The AUC scores for different fractions of removed links for three networks: \textbf{(a)} the undirected and weighted social network \texttt{moviegalaxies/34}, \textbf{(b)} the undirected and unweighted informational network \texttt{wiki\_science}, and \textbf{(c)} the directed and unweighted biological network \texttt{ecoli\_transcription/v1.0}. The $95\%$ confidence intervals are so narrow that they are not visible in the plot.}
    \label{fig:single_networks_1}
\end{figure*}

\subsection{Three example networks}
\noindent%
To illustrate the performance of the link prediction methods, we examine three example networks more closely: the undirected and weighted social network \texttt{moviegalaxies/34} (\cref{fig:single_networks_1}a), the undirected and unweighted informational network \texttt{wiki\_science} (\cref{fig:single_networks_1}b), and the directed and unweighted biological network \texttt{ecoli\_transcription/v1.0} (\cref{fig:single_networks_1}c).
Overall, standard and globally regularized MapSim perform similarly on the social network, with a slight advantage for standard MapSim. However, as the fraction of removed links increases, the performance of standard MapSim declines --- a trend common across most investigated networks.
For the informational network, the two methods perform comparably up to approximately $50\%$ of removed links.
Beyond this point, the performance of MapSim with global regularization drops significantly because the regularized map equation determines that the available data are insufficient to infer reliable community structure, assigning all nodes to a single community.
In this case, standard MapSim's link prediction based on unreliable modules has a limited effect on its performance. 
In other cases, such as the biological network in \cref{fig:single_networks_1}c, the unreliable modules impair standard MapSim's performance, and MapSim with global regularization produces better results.
This variability shows that no single method consistently performs best across all networks~\cite{peel2017ground}.

\begin{figure*}[htp!]
    \centering    \includegraphics[width=0.97\textwidth]{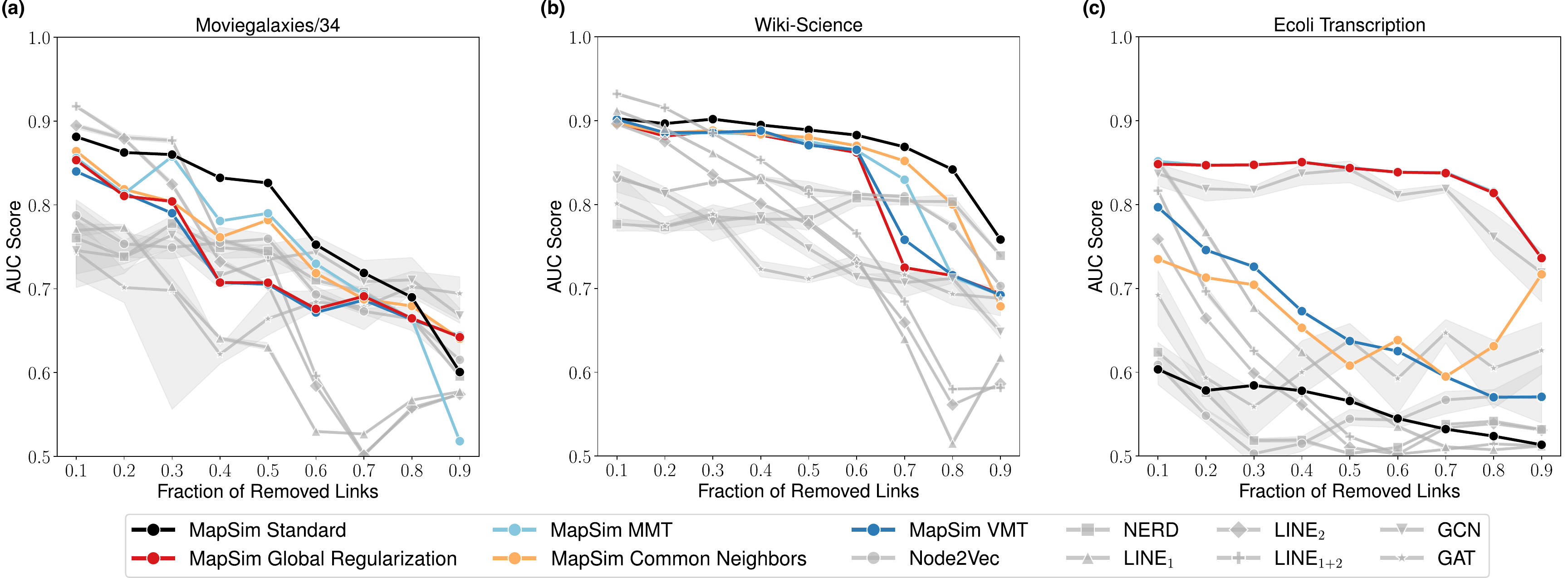}
    \caption{\textbf{The AUC Scores for different link prediction methods.} This example displays three networks. \textbf{(a)} is the social network \texttt{moviegalaxies/34} which is undirected and weighted. \textbf{(b)} is the informational network \texttt{wiki\_science} which is undirected and unweighted. Finally, \textbf{(c)} represents the biological network \texttt{ecoli\_transcription/v1.0} which is directed and unweighted.}
    \label{fig:single_networks_3}
\end{figure*}

To address the decline of MapSim with global regularization as networks become too sparse, we combined global and local regularization (\cref{eq:weight_sum}).
These methods integrate global regularization for robust community detection with local adjustments to possibly improve prediction performance.
We compared the performance on all example networks of all versions of MapSim together with the seven state-of-the-art embedding methods (\cref{fig:single_networks_3}).
In the social network, all MapSim versions performed similarly, with standard MapSim outperforming the others until a large fraction of links has been removed (\cref{fig:single_networks_3}a). MapSim consistently outperforms all embedding methods.
In the informational network, the local regularization methods demonstrate improved prediction performance even for large fractions of removed links, where MapSim with global regularization struggles to find community structures (\cref{fig:single_networks_3}b).
The local regularization methods also drop in performance but at higher fractions of removed links.
LINE$_{1+2}$ and LINE$_2$ exhibit strong performance initially but falter as the fraction of removed links increases.
In the biological network, MMT performs comparably to MapSim with global regularization, while VMT and Common Neighbors show lower performance but still outperform standard MapSim.
The embedding methods yield results similar to standard MapSim, with GCN coming close to the performance of MMT and MapSim with global regularization (\cref{fig:single_networks_3}c).
Again, no single method performs best across all networks.  

\subsection{Average performance}
\begin{figure*}[htp!]
    \centering
    \includegraphics[width=1.0\textwidth]{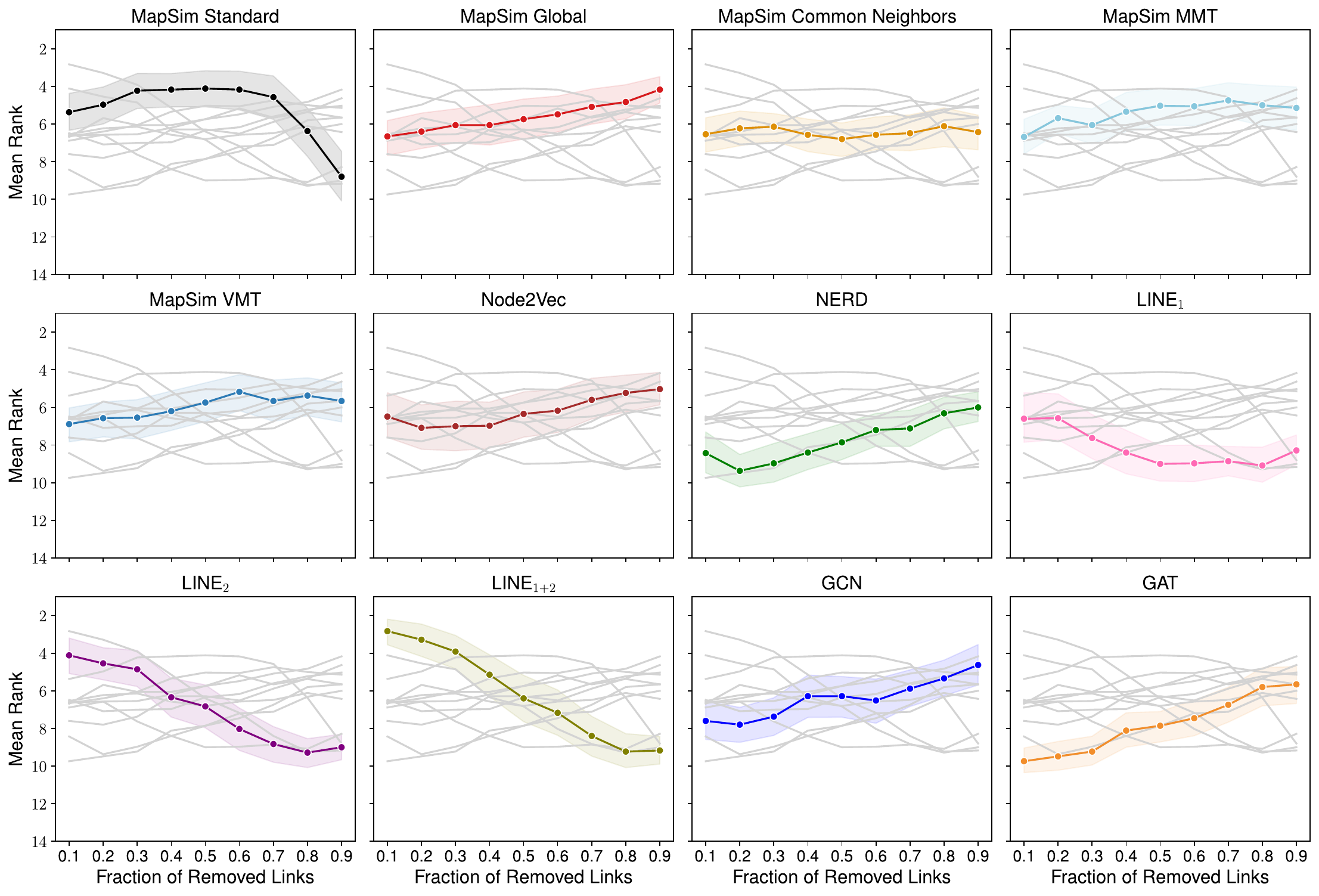}
    \caption{\textbf{The Mean Rank of different link prediction methods.} In each figure, the named method is depicted with a $95\%$ bootstrapped confidence interval. The other methods are included for comparison with light grey lines in the background.}
    \label{fig:mean_rank}
\end{figure*}

\noindent%
To extend the analysis beyond the three example networks, we evaluated the methods on 35 networks.
We computed the rank of each method for every network, with rank 1 denoting the top-performing method based on the AUC score.
We then calculated the mean rank and computed a $95\%$ confidence interval using percentile bootstrapping~\cite{bootstrap}.
We randomly sampled 1,000 bootstrap samples with replacement from the original data and calculated the upper and lower percentiles from the bootstrap distribution of ranks (\cref{fig:mean_rank}; corresponding AUC scores in \cref{fig:mean_auc}, and tabulated AUC scores and ranks in \Cref{tab:mean_auc} and \cref{tab:mean_rank}, respectively).

We found that standard MapSim has good link prediction performance in dense networks but has a drop in performance when the fraction of removed links becomes high.
In contrast, MapSim with global regularization maintains stable performance even for very sparse networks. 
The methods incorporating local regularization also show a stable performance and perform almost on par with globally regularized MapSim. While LINE$_{1+2}$ and LINE$_2$ excel in denser networks and the GNN method GCN performs well in sparse networks, MapSim with global regularization consistently outperforms the embedding methods across most scenarios.

Determining the sparsity of a real-world network can be challenging when its actual density is unknown. To guide the choice between standard MapSim and MapSim with global regularization, we suggest the following approach: First, leverage domain expertise. Experts familiar with the network's characteristics can often make an informed estimate of its expected density, guiding the choice of method. Second, perform structural analysis. When direct density estimation is difficult, examining the presence and stability of community structures after removing a fraction of links can indicate whether the network is dense or sparse. Third, choose a conservative approach. When uncertainty remains, MapSim with global regularization is generally the safer choice. While standard MapSim performs slightly better in denser networks, MapSim with global regularization maintains strong performance across a range of densities and significantly outperforms standard MapSim in sparse networks.

\subsection{Community detection}
\begin{figure*}[htp!]
    \centering
    \includegraphics[width=1.0\textwidth]{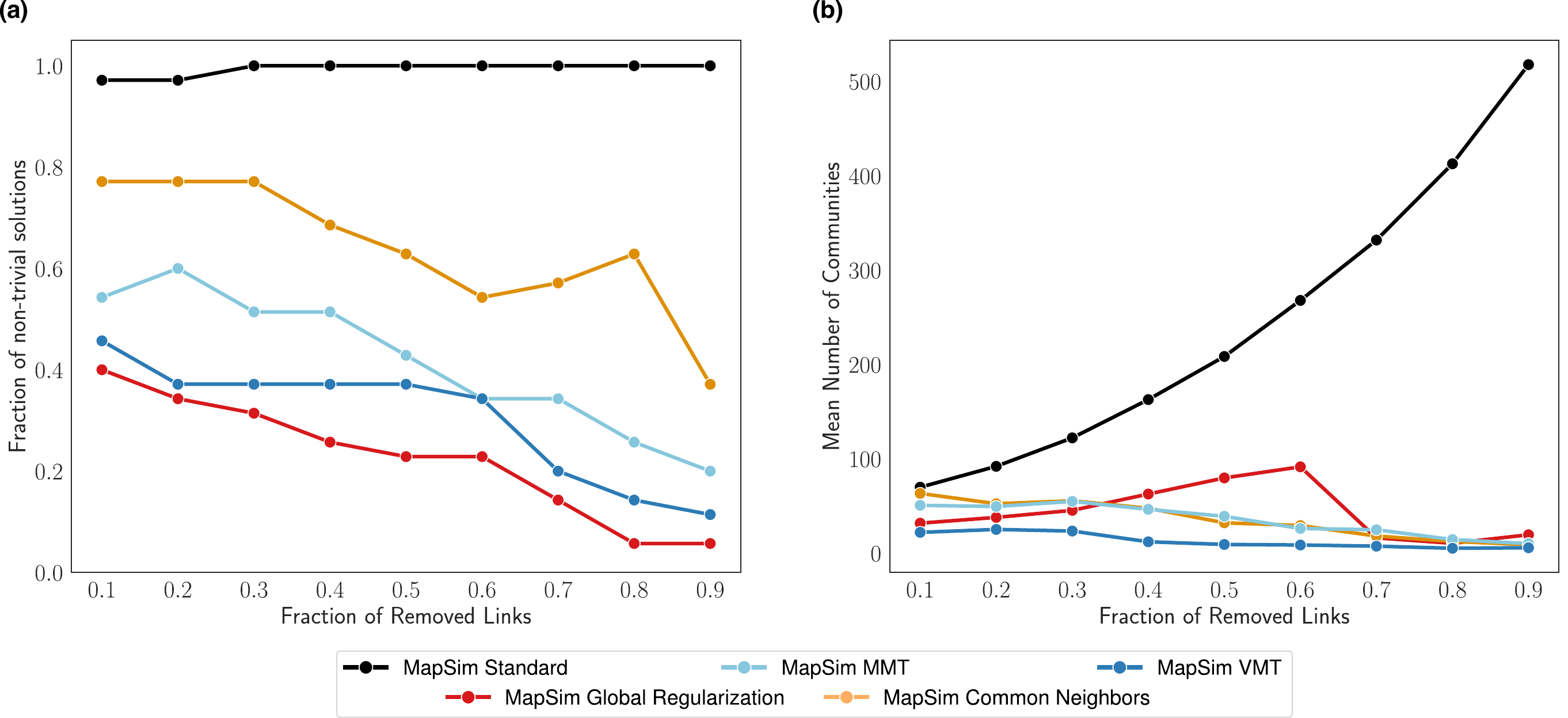}
    \caption{\textbf{Community detection.} \textbf{(a)} The fraction of non-trivial solution for the different map equation methods. \textbf{(b)} The mean number of communities detected by each map equation method.}
    \label{fig:non_triv_communities}
\end{figure*}

\noindent%
MapSim achieves reliable link prediction when it accurately detects the communities within the network.
In sparse networks, the standard map equation can suffer from over-partitioning, resulting in an increasing number of communities as the network becomes sparser.
Conversely, when the globally regularized map equation lacks sufficient information to determine the community structure, it tends to collapse into a single module.
To evaluate this effect on the investigated networks, we computed the fraction of non-trivial solutions for each fraction of removed links, where a trivial solution is defined as having only one community.
From these non-trivial solutions, we calculated the mean number of communities for each fraction of removed links (\cref{fig:non_triv_communities}).
While the standard map equation almost exclusively prefers non-trivial solutions, the regularized versions often collapse the network into one community.
The regularized methods handle sparse networks better, resulting in improved link prediction performance.

\begin{figure*}[htp!]
    \centering
    \includegraphics[width=1\textwidth]{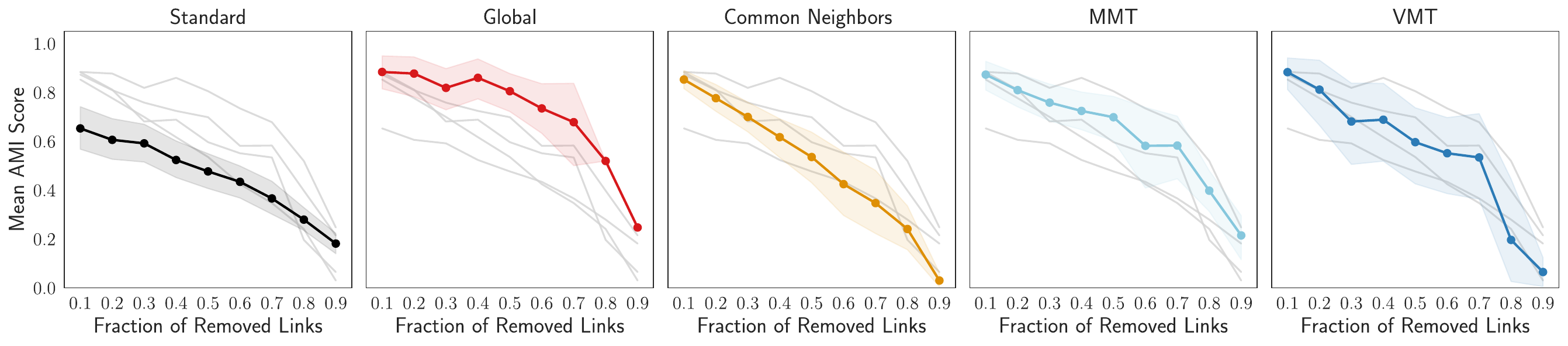}
  \caption{\textbf{The Mean AMI Scores of non-trivial solutions.} Each figure highlights the named map equation method with a $95\%$ bootstrapped confidence interval, while other methods are shown as light grey lines in the background for comparison.}
  \label{fig:mean_ami_scores}
\end{figure*}

Since no ground truth communities exist, we estimated them by applying the map equation to each network before removing any links.
We then compared these calculated ``true'' communities with the non-trivial communities detected for each fraction of removed links and calculated the mean Adjusted Mutual Information (AMI) scores~\cite{ami}.
These results are illustrated in \cref{fig:mean_ami_scores}, where the highlighted method is depicted with a $95\%$ bootstrapped confidence interval.
These findings indicate that community detection using the map equation with global regularization consistently outperforms the local regularizations as well as the standard map equation, regardless of the fraction of removed links.
The standard map equation has the lowest performance, while the local regularization methods perform better but still fall short of the globally regularized map equation.

\subsection{Scalability}
\noindent%
To investigate the scalability of the best-performing method on sparse networks, MapSim with global regularization, we extended our evaluation to networks roughly an order of magnitude larger than the largest previously analyzed. In these cases, we compared only with the GNN methods and standard MapSim, excluding other embedding methods due to their extensive hyperparameter tuning and prohibitive training times. For example, node2vec was more than 1,000 times slower than MapSim with global regularization (\cref{tab:mean_time}). With no need for hyperparameter tuning, MapSim remains significantly faster.

\begin{figure*}[htp!]
    \centering
    \includegraphics[width=1.0\textwidth]{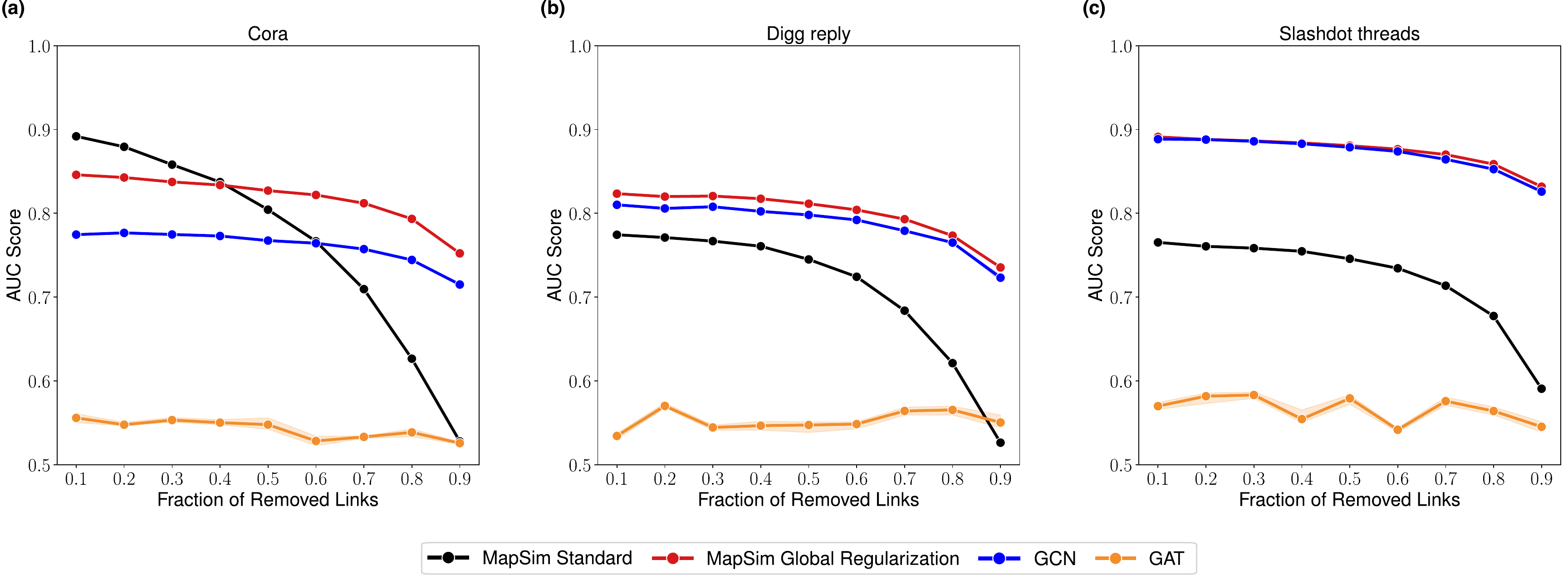}
    \caption{\textbf{Link prediction performance on three larger networks.} \textbf{(a)} is the informational network \texttt{cora}, \textbf{(b)} is the social network \texttt{digg\_reply}, and \textbf{(c)} represents the communication network \texttt{slashdot\_threads}.}
    \label{fig:large_networks}
\end{figure*}

\cref{fig:large_networks} presents the AUC scores on three larger networks (\cref{tab:dataset_summary}): the informational network \texttt{cora} (\cref{fig:large_networks}a), the social network \texttt{digg\_reply} (\cref{fig:large_networks}b), and the communication network \texttt{slashdot\_threads} (\cref{fig:large_networks}c). Aside from GAT’s underperformance, the trends in these results align with previous findings --- standard MapSim performs well in dense networks but declines as sparsity increases. Like the hyperparameter-tuned GCN, MapSim with global regularization remains more stable.

\section{Conclusion}
\noindent%
We regularize the map equation in various ways to enhance link prediction with MapSim.
Our findings demonstrate that global regularization consistently outperforms standard MapSim and other state-of-the-art embedding methods in highly sparse networks.
Although standard MapSim excels in denser networks, global regularization maintains stable performance also in sparse networks, making it a good choice when the actual network density is unknown.
Global regularization also remains reliable by preventing over-partitioning in sparse networks.

While global regularization offers the most consistent overall performance, local regularization methods, such as Common Neighbors, Mixed Markov Time, and Variable Markov Time, perform best in certain networks. 

In summary, while standard MapSim is effective in dense networks, many real-world networks have unknown actual density due to unobserved or missing links. In such cases, MapSim with global regularization offers a more robust and reliable alternative.
This approach ensures reliable community detection, high interpretability, and robust link prediction across diverse networks --- all without the need for hyperparameter tuning and with significantly lower computational cost than embedding-based methods --- making it a valuable tool for improving link prediction accuracy in a wide array of applications.

\section{Declarations}

\subsection{Availability of data and materials}
\noindent%
The networks were retrieved from \url{https://networks.skewed.de}. The source code for MapSim is freely available online at \url{https://github.com/mapequation/map-equation-similarity}. The source code for NERD can be found at \url{https://git.l3s.uni-hannover.de/khosla/nerd}.
The source code for LINE is available at \url{https://github.com/tangjianpku/LINE}.

\subsection{Competing interests}
\noindent%
M.L.\ is employed by Siftlab AB, which owns Infomap's commercial rights, and M.R.\ is a co-founder and a board member of Siftlab AB. The other authors declare no competing interests.

\subsection{Funding}
\noindent%
M.L.\ was supported by the Wallenberg AI, Autonomous Systems and Software Program (\url{http://wasp-sweden.org}), funded by the Knut and Alice Wallenberg Foundation. M.R. was supported by the Swedish Research Council, grant 2023-03705.
C.B.\ was supported by the Swiss National Science Foundation, grant no.\ 176938.

\subsection{Authors' contributions}
\noindent%
M.L.\ and C.B.\ performed the experiments. All authors wrote, edited, and accepted the manuscript in its final form.

\subsection{Acknowledgements}
\noindent%
We thank Daniel Edler for helpful discussions and help with the implementation of Variable Markov Time. The computations were enabled by resources provided by the Swedish National Infrastructure for Computing at the High Performance Computer Center North (HPC2N) in Umeå, Sweden, partially funded by the Swedish Research Council through grant agreement no.~2018-05973, and by the National Academic Infrastructure for Supercomputing in Sweden (NAISS), partially funded by the Swedish Research Council through grant agreement no.~2022-06725.

\bibliographystyle{unsrt}

\onecolumngrid
\appendix
\renewcommand{\thefigure}{A\arabic{figure}}
\renewcommand{\thetable}{A\Roman{table}}
\setcounter{figure}{0}

\clearpage
\section*{Appendix}

\begin{figure*}[h!]\centering\includegraphics[width=1.0\textwidth]{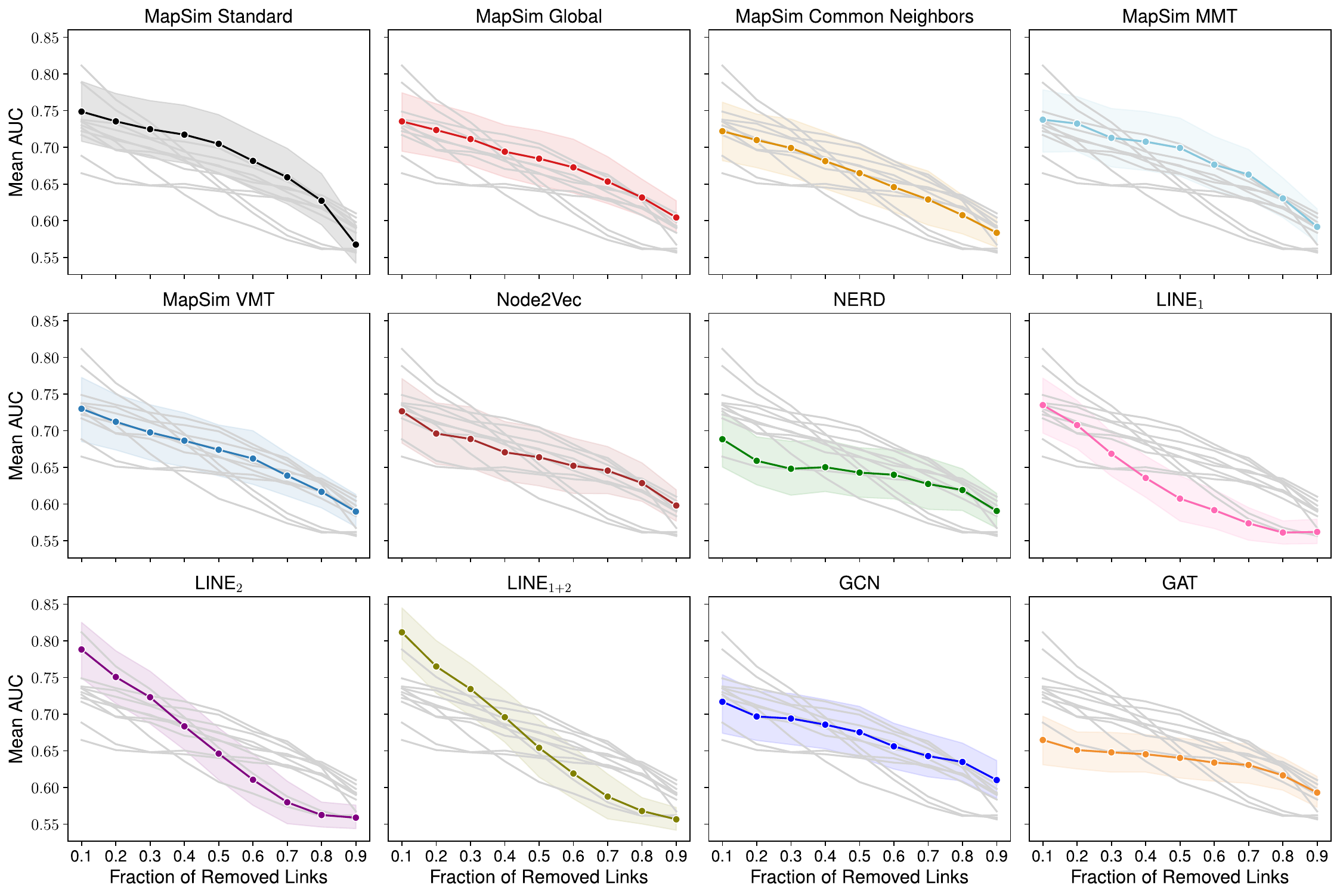}
    \caption{\textbf{The Mean AUC for different link prediction methods for each fraction of removed links.} In each figure, the method that is highlighted shows a $95\%$ bootstrapped confidence interval calculated. The other methods can be seen as light grey lines in the background.}
    \label{fig:mean_auc}
\end{figure*}

\begin{table*}[h!]
    \centering
    \caption{\textbf{Summary of the 38 real-world networks used.} The three networks with stars were not included in the average performance analysis, only in the scalability analysis.}
\begin{tabular}{lcrrcc}
        \toprule
        \textbf{Networks} & Ref & Nodes & Edges & Weighted & Directed \\
        \midrule
        adjnoun & \cite{adjnoun} & 112 & 425 & No & No \\
        bison & \cite{bison} & 26 & 314 & Yes & Yes \\
        blumenau\_drug & \cite{blumenau_drug} & 75 & 181 & Yes & No \\
        cattle & \cite{cattle} & 28 & 217 & Yes & Yes \\
        celegans\_metabolic & \cite{celegans_metabolic} & 453 & 4,596 & No & No \\
        collins\_yeast & \cite{collins_yeast} & 1,622 & 9,070 & No & No \\
        dolphins & \cite{dolphins} & 62 & 159 & No & No \\
\bottomrule
    \end{tabular}
    \label{tab:dataset_summary}
\end{table*}

\begin{table*}[h!]
    \centering 
    \caption{\textbf{The Mean AUC scores of MapSim MMT for different values of $\beta$ and for each fraction of removed links.} For each fraction of removed links, the best AUC score is shown in bold and the second best is underlined. Two significant
    digits are shown but the evaluation was done on three significant digits.}
    \adjustbox{max width=\textwidth}{
    \begin{tabular}{l*{9}{c}}
        \toprule
        & {$\beta=0.1$} & {$\beta=0.2$} & {$\beta=0.3$} & {$\beta=0.4$} & {$\beta=0.5$} & {$\beta=0.6$} & {$\beta=0.7$} & {$\beta=0.8$} & {$\beta=0.9$} \\
        \midrule
        0.1 & 0.73 & \textbf{0.73} & \textbf{0.73} & 0.73 & 0.73 & 0.73 & 0.73 & 0.73 & 0.73 \\
        & {\textcolor{gray}{(0.68, 0.77)}} & {\textcolor{gray}{(0.69, 0.77)}} & {\textcolor{gray}{(0.69, 0.77)}} & {\textcolor{gray}{(0.68, 0.77)}} & {\textcolor{gray}{(0.68, 0.77)}} & {\textcolor{gray}{(0.68, 0.77)}} & {\textcolor{gray}{(0.69, 0.77)}} & {\textcolor{gray}{(0.68, 0.77)}} & {\textcolor{gray}{(0.69, 0.77)}} \\
        0.2 & \textbf{0.73} & 0.73 & 0.73 & \textbf{0.73} & 0.73 & 0.73 & 0.73 & 0.73 & 0.73 \\
        & {\textcolor{gray}{(0.69, 0.78)}} & {\textcolor{gray}{(0.69, 0.77)}} & {\textcolor{gray}{(0.69, 0.77)}} & {\textcolor{gray}{(0.69, 0.77)}} & {\textcolor{gray}{(0.69, 0.77)}} & {\textcolor{gray}{(0.69, 0.77)}} & {\textcolor{gray}{(0.69, 0.77)}} & {\textcolor{gray}{(0.69, 0.77)}} & {\textcolor{gray}{(0.69, 0.77)}} \\
        0.3 & \textbf{0.72} & \textbf{0.72} & 0.72 & 0.71 & 0.72 & 0.71 & 0.71 & 0.72 & 0.72 \\
        & {\textcolor{gray}{(0.68, 0.76)}} & {\textcolor{gray}{(0.68, 0.76)}} & {\textcolor{gray}{(0.68, 0.76)}} & {\textcolor{gray}{(0.68, 0.75)}} & {\textcolor{gray}{(0.68, 0.76)}} & {\textcolor{gray}{(0.68, 0.76)}} & {\textcolor{gray}{(0.68, 0.76)}} & {\textcolor{gray}{(0.68, 0.76)}} & {\textcolor{gray}{(0.68, 0.76)}} \\
        0.4 & 0.71 & 0.71 & \underline{0.71} & \textbf{0.71} & \underline{0.71} & 0.71 & 0.71 & 0.71 & 0.71 \\
\bottomrule
    \end{tabular}
    }
    \label{tab:beta_MMT}    
\end{table*}

\begin{table*}[h!]
    \centering 
    \caption{\textbf{The Mean AUC scores for different link prediction methods for each fraction of removed links.} For each fraction of removed links, the best AUC score is shown in bold and the second best is underlined. Two significant digits are shown but the evaluation was done on three significant digits. In the few cases where AUC $< 0.5$, we flipped it by ($1 -$AUC) $> 0.5$.}
    \adjustbox{max width=\textwidth}{%
    \begin{tabular}{l*{12}{c}}
        \toprule
        & {MapSim} & {MapSim} & {Common} & {MapSim} & {MapSim} & {node2Vec} & {NERD} & {LINE$_1$} & {LINE$_2$} & {LINE$_{1+2}$} & {GCN} & {GAT} \\
        & {Standard} & {Global} & {Neighbors} & {MMT} & {VMT} & & & & & \\
        \midrule
        0.1 & 0.75 & 0.74 & 0.72 & 0.74 & 0.73 & 0.73 & 0.69 & 0.73 & \underline{0.79} & \textbf{0.81} & 0.72 & 0.66 \\
        & {\textcolor{gray}{(0.71, 0.79)}} & {\textcolor{gray}{(0.69, 0.77)}} & {\textcolor{gray}{(0.68, 0.76)}} & {\textcolor{gray}{(0.69, 0.78)}} & {\textcolor{gray}{(0.69, 0.77)}} & {\textcolor{gray}{(0.68, 0.77)}} & {\textcolor{gray}{(0.65, 0.73)}} & {\textcolor{gray}{(0.70, 0.77)}} & {\textcolor{gray}{(0.75, 0.83)}} & {\textcolor{gray}{(0.78, 0.85)}} & {\textcolor{gray}{(0.67, 0.75)}} & {\textcolor{gray}{(0.63, 0.70)}} \\
        0.2 & 0.74 & 0.72 & 0.71 & 0.73 & 0.71 & 0.70 & 0.66 & 0.71 & \underline{0.75} & \textbf{0.77} & 0.70 & 0.65 \\
        & {\textcolor{gray}{(0.70, 0.77)}} & {\textcolor{gray}{(0.69, 0.76)}} & {\textcolor{gray}{(0.67, 0.75)}} & {\textcolor{gray}{(0.69, 0.77)}} & {\textcolor{gray}{(0.67, 0.75)}} & {\textcolor{gray}{(0.65, 0.74)}} & {\textcolor{gray}{(0.63, 0.69)}} & {\textcolor{gray}{(0.68, 0.74)}} & {\textcolor{gray}{(0.71, 0.79)}} & {\textcolor{gray}{(0.73, 0.80)}} & {\textcolor{gray}{(0.66, 0.73)}} & {\textcolor{gray}{(0.63, 0.68)}} \\
        0.3 & \underline{0.72} & 0.71 & 0.70 & 0.71 & 0.70 & 0.69 & 0.65 & 0.67 & 0.72 & \textbf{0.73} & 0.69 & 0.65 \\
        & {\textcolor{gray}{(0.69, 0.76)}} & {\textcolor{gray}{(0.68, 0.75)}} & {\textcolor{gray}{(0.66, 0.74)}} & {\textcolor{gray}{(0.67, 0.75)}} & {\textcolor{gray}{(0.66, 0.74)}} & {\textcolor{gray}{(0.65, 0.73)}} & {\textcolor{gray}{(0.61, 0.69)}} & {\textcolor{gray}{(0.64, 0.70)}} & {\textcolor{gray}{(0.69, 0.76)}} & {\textcolor{gray}{(0.70, 0.77)}} & {\textcolor{gray}{(0.66, 0.73)}} & {\textcolor{gray}{(0.62, 0.68)}} \\
\bottomrule
    \end{tabular}
    }
    \label{tab:mean_auc}
\end{table*}

\begin{table*}[h!]
    \caption{\textbf{The Mean rank for different link prediction methods for each fraction of removed links.} For each fraction of removed links, the best rank is shown in bold and the second best is underlined. Two significant digits are shown but the evaluation was done on three significant digits.}
    \adjustbox{max width=\textwidth}{%
    \begin{tabular}{l*{12}{c}}
        \toprule
        & {MapSim} & {MapSim} & {Common} & {MapSim} & {MapSim} & {node2Vec} & {NERD} & {LINE$_1$} & {LINE$_2$} & {LINE$_{1+2}$} & {GCN} & {GAT} \\
        & {Standard} & {Global} & {Neighbors} & {MMT} & {VMT} & & & & & \\
        \midrule
        0.1 & 5.4 & 6.7 & 6.5 & 6.7 & 6.9 & 6.5 & 8.4 & 6.6 & \underline{4.1} & \textbf{2.8} & 7.6 & 9.7 \\
        & {\textcolor{gray}{(4.4, 6.3)}} & {\textcolor{gray}{(5.8, 7.7)}} & {\textcolor{gray}{(5.7, 7.5)}} & {\textcolor{gray}{(5.8, 7.6)}} & {\textcolor{gray}{(6.0, 7.8)}} & {\textcolor{gray}{(5.2, 7.6)}} & {\textcolor{gray}{(7.3, 9.5)}} & {\textcolor{gray}{(5.2, 7.9)}} & {\textcolor{gray}{(3.2, 5.1)}} & {\textcolor{gray}{(2.2, 3.6)}} & {\textcolor{gray}{(6.5, 8.6)}} & {\textcolor{gray}{(9.0, 10.3)}} \\
        0.2 & 5.0 & 6.4 & 6.2 & 5.7 & 6.6 & 7.1 & 9.4 & 6.6 & \underline{4.5} & \textbf{3.3} & 7.8 & 9.5 \\
        & {\textcolor{gray}{(4.0, 6.0)}} & {\textcolor{gray}{(5.4, 7.3)}} & {\textcolor{gray}{(5.3, 7.2)}} & {\textcolor{gray}{(5.0, 6.5)}} & {\textcolor{gray}{(5.7, 7.6)}} & {\textcolor{gray}{(5.9, 8.2)}} & {\textcolor{gray}{(8.5, 10.2)}} & {\textcolor{gray}{(5.3, 7.7)}} & {\textcolor{gray}{(3.7, 5.4)}} & {\textcolor{gray}{(2.5, 4.2)}} & {\textcolor{gray}{(6.9, 8.7)}} & {\textcolor{gray}{(8.7, 10.2)}} \\
        0.3 & \underline{4.2} & 6.1 & 6.1 & 6.1 & 6.5 & 7.0 & 9.0 & 7.6 & 4.9 & \textbf{3.9} & 7.4 & 9.2 \\
        & {\textcolor{gray}{(3.3, 5.2)}} & {\textcolor{gray}{(5.2, 7.0)}} & {\textcolor{gray}{(5.4, 6.9)}} & {\textcolor{gray}{(5.2, 7.1)}} & {\textcolor{gray}{(5.6, 7.7)}} & {\textcolor{gray}{(5.7, 8.3)}} & {\textcolor{gray}{(7.9, 10.0)}} & {\textcolor{gray}{(6.4, 8.7)}} & {\textcolor{gray}{(3.9, 5.8)}} & {\textcolor{gray}{(3.1, 4.8)}} & {\textcolor{gray}{(6.4, 8.4)}} & {\textcolor{gray}{(8.4, 9.9)}} \\
\end{tabular}
    }
    
\end{table*}

\begin{table*}
    \centering 
        \caption{\textbf{The average execution time in seconds over all fraction of removed links for each network and method.}}
    \adjustbox{max width=\textwidth}{%
    \begin{tabular}{l*{12}{r}}
    \toprule
     & {Node2Vec} & {NERD} & {LINE$_1$} & {LINE$_2$} & {LINE$_{1+2}$} & {MapSim} & {MapSim} & {Common} & {MapSim} & {MapSim} & {GCN} & {GAT} \\
    & & & & & & {Standard} & {Global} & {Neighbors} & {MMT} & {VMT} &  \\
    \midrule
    adjnoun & 615.0 & 450.4 & 65.2 & 61.7 & 112.5 & 0.9 & 0.2 & 1.7 & 1.1 & 1.1 & 5.7 & 4.9 \\
    bison & 128.9 & 738.4 & 85.4 & 72.6 & 142.6 & 0.2 & 0.0 & 0.1 & 0.1 & 0.2 & 4.3 & 3.3 \\
    blumenau\_drug & 386.5 & 753.3 & 79.9 & 82.8 & 128.2 & 0.6 & 0.1 & 1.7 & 1.7 & 0.7 & 4.4 & 4.0 \\
    cattle & 114.9 & 899.0 & 136.5 & 113.7 & 193.9 & 0.4 & 0.1 & 0.1 & 0.1 & 0.2 & 2.4 & 3.3 \\
    celegans\_metabolic & 5927.0 & 370.2 & 59.7 & 59.4 & 97.9 & 3.4 & 1.0 & 65.5 & 56.3 & 26.0 & 19.9 & 16.1 \\
    collins\_yeast & 24107.0 & 306.6 & 64.2 & 59.8 & 110.4 & 16.7 & 6.2 & 576.5 & 586.7 & 731.4 & 76.1 & 45.7 \\
\end{tabular}}
\end{table*}

\clearpage
\begin{longtable}{lrrrrrrrrrr}
    \caption{\textbf{The total link weights of the networks.} The total link weight, or density, of each network after applying different MapSim regularization methods for each fraction of removed links. Original is the network without any regularization.} \label{tab:density} \\

    \toprule
    Network & & 0.1 & 0.2 & 0.3 & 0.4 & 0.5 & 0.6 & 0.7 & 0.8 & 0.9 \\
    \midrule
    \endfirsthead
    
    \toprule
    Network & & 0.1 & 0.2 & 0.3 & 0.4 & 0.5 & 0.6 & 0.7 & 0.8 & 0.9 \\
    \midrule
    \endhead
    
    \midrule
    \multicolumn{11}{r}{\textit{Continued on next page}} \\
    \midrule
    \endfoot
    
    \bottomrule
    \endlastfoot

    adjnoun & Original & 382 & 340 & 298 & 255 & 212 & 170 & 128 & 85 & 42 \\
     & Global & 772 & 730 & 688 & 645 & 602 & 560 & 518 & 475 & 432 \\
     & MMT & 1233 & 1147 & 1061 & 972 & 885 & 800 & 708 & 613 & 510 \\
     & Common Neighbors & 1761 & 1612 & 1449 & 1323 & 1171 & 1024 & 885 & 713 & 546 \\
     & VMT & 1154 & 1026 & 986 & 866 & 783 & 730 & 646 & 560 & 465 \\\midrule
    bison & Original & 811 & 717 & 607 & 520 & 457 & 361 & 264 & 172 & 100 \\
     & Global & 848 & 754 & 644 & 557 & 494 & 398 & 301 & 209 & 137 \\
     & MMT & 1136 & 1045 & 932 & 834 & 802 & 691 & 602 & 570 & 422 \\
\end{longtable}

\begin{longtable}{lrrrrrrrrrr}
    \caption{\textbf{Optimal Node2Vec Hyperparameters.} For each fraction of removed links and each network we show the optimal hyperparameters used for the results. We conducted $r \in \left\{60, 70, 80, 90, 100\right\}$ random walks of length $l \in \left\{30, 40, 50, 60\right\}$ per node and employed a window size of $w \in \left\{5, 10, 15\right\}$.
We set the in-out parameter $p$, and the return parameter $q$, as $p,q \in \left\{0.5, 1, 2, 4\right\}$. The embeddings were configured with $d \in \left\{64, 128, 256\right\}$ dimensions.} \label{tab:hyperparams_Node2Vec} \\

    \toprule
    Network & & 0.1 & 0.2 & 0.3 & 0.4 & 0.5 & 0.6 & 0.7 & 0.8 & 0.9 \\
    \midrule
    \endfirsthead
    
    \toprule
    Network & & 0.1 & 0.2 & 0.3 & 0.4 & 0.5 & 0.6 & 0.7 & 0.8 & 0.9 \\
    \midrule
    \endhead
    
    \midrule
    \multicolumn{11}{r}{\textit{Continued on next page}} \\
    \midrule
    \endfoot
    
    \bottomrule
    \endlastfoot

    adjnoun & $r$ & 90 & 80 & 60 & 80 & 80 & 90 & 90 & 100 & 80 \\
     & $p$ & 4 & 1 & 1 & 4 & 4 & 4 & 4 & 4 & 0 \\
     & $q$ & 2 & 0 & 0 & 0 & 0 & 0 & 0 & 2 & 1 \\
     & $d$ & 128 & 128 & 128 & 128 & 256 & 128 & 128 & 128 & 256 \\
     & $l$ & 40 & 50 & 50 & 60 & 50 & 60 & 60 & 30 & 40 \\
     & $w$ & 10 & 10 & 10 & 15 & 10 & 10 & 10 & 15 & 10 \\\midrule
    bison & $r$ & 90 & 80 & 80 & 100 & 90 & 100 & 60 & 60 & 100 \\
     & $p$ & 1 & 1 & 2 & 2 & 0 & 4 & 4 & 4 & 1 \\
\end{longtable}

\begin{longtable}{lrrrrrrrrrr}
    \caption{\textbf{Optimal NERD Hyperparameters.} For each fraction of removed links and each network we show the optimal hyperparameters used for the results. We created $s \in \left\{2, 3, 4, 5\right\}$ negative samples and set the starting value of the learning rate to $\rho \in \left\{0.0001, 0.001, 0.01\right\}$.
 The walk size was set to $n \in \left\{1, 2, 3, 4, 5\right\}$ and the embeddings were configured with $d \in \left\{64, 128, 256\right\}$ dimensions.} \label{tab:hyperparams_NERD} \\

    \toprule
    Network & & 0.1 & 0.2 & 0.3 & 0.4 & 0.5 & 0.6 & 0.7 & 0.8 & 0.9 \\
    \midrule
    \endfirsthead
    
    \toprule
    Network & & 0.1 & 0.2 & 0.3 & 0.4 & 0.5 & 0.6 & 0.7 & 0.8 & 0.9 \\
    \midrule
    \endhead
    
    \midrule
    \multicolumn{11}{r}{\textit{Continued on next page}} \\
    \midrule
    \endfoot
    
    \bottomrule
    \endlastfoot

    adjnoun & $s$ & 4 & 5 & 5 & 4 & 5 & 5 & 5 & 5 & 4 \\
     & $d$ & 128 & 64 & 64 & 128 & 128 & 64 & 256 & 128 & 64 \\
     & $\rho$ & 0.01 & 0.01 & 0.001 & 0.01 & 0.01 & 0.01 & 0.01 & 0.01 & 0.01 \\
     & $n$ & 2 & 5 & 5 & 4 & 4 & 2 & 3 & 5 & 5 \\\midrule
    bison & $s$ & 4 & 3 & 5 & 2 & 2 & 3 & 3 & 4 & 5 \\
     & $d$ & 64 & 64 & 64 & 128 & 128 & 128 & 256 & 64 & 64 \\
     & $\rho$ & 0.001 & 0.01 & 0.01 & 0.001 & 0.001 & 0.01 & 0.01 & 0.01 & 0.0001 \\
     & $n$ & 3 & 5 & 2 & 1 & 4 & 2 & 3 & 4 & 5 \\
\end{longtable}

\begin{longtable}{lrrrrrrrrrr}
    \caption{\textbf{Optimal LINE$_1$ Hyperparameters.} For each fraction of removed links and each network we show the optimal hyperparameters used for the results. We created $s \in \left\{2, 3, 4, 5\right\}$ negative samples and set the starting value of the learning rate to $\rho \in \left\{0.0001, 0.001, 0.01\right\}$.
The embeddings were configured with $d \in \left\{64, 128, 256\right\}$ dimensions.} \label{tab:hyperparams_LINE1} \\

    \toprule
    Network & & 0.1 & 0.2 & 0.3 & 0.4 & 0.5 & 0.6 & 0.7 & 0.8 & 0.9 \\
    \midrule
    \endfirsthead
    
    \toprule
    Network & & 0.1 & 0.2 & 0.3 & 0.4 & 0.5 & 0.6 & 0.7 & 0.8 & 0.9 \\
    \midrule
    \endhead
    
    \midrule
    \multicolumn{11}{r}{\textit{Continued on next page}} \\
    \midrule
    \endfoot
    
    \bottomrule
    \endlastfoot

    adjnoun & $s$ & 5 & 3 & 2 & 3 & 5 & 3 & 3 & 2 & 5 \\
     & $d$ & 64 & 128 & 128 & 64 & 256 & 128 & 256 & 256 & 256 \\
     & $\rho$ & 0.001 & 0.001 & 0.01 & 0.001 & 0.0001 & 0.0001 & 0.0001 & 0.0001 & 0.0001 \\\midrule
    bison & $s$ & 4 & 3 & 3 & 2 & 3 & 3 & 2 & 4 & 2 \\
     & $d$ & 128 & 64 & 256 & 256 & 256 & 128 & 256 & 256 & 128 \\
     & $\rho$ & 0.0001 & 0.0001 & 0.0001 & 0.0001 & 0.0001 & 0.0001 & 0.0001 & 0.0001 & 0.001 \\\midrule
    blumenau\_drug & $s$ & 3 & 2 & 2 & 5 & 2 & 2 & 3 & 5 & 2 \\
     & $d$ & 256 & 256 & 64 & 256 & 256 & 64 & 128 & 128 & 128 \\
\end{longtable}

\begin{longtable}{lrrrrrrrrrr}
    \caption{\textbf{Optimal LINE$_2$ Hyperparameters.} For each fraction of removed links and each network we show the optimal hyperparameters used for the results. We created $s \in \left\{2, 3, 4, 5\right\}$ negative samples and set the starting value of the learning rate to $\rho \in \left\{0.0001, 0.001, 0.01\right\}$.
The embeddings were configured with $d \in \left\{64, 128, 256\right\}$ dimensions.} \label{tab:hyperparams_LINE2} \\

    \toprule
    Network & & 0.1 & 0.2 & 0.3 & 0.4 & 0.5 & 0.6 & 0.7 & 0.8 & 0.9 \\
    \midrule
    \endfirsthead
    
    \toprule
    Network & & 0.1 & 0.2 & 0.3 & 0.4 & 0.5 & 0.6 & 0.7 & 0.8 & 0.9 \\
    \midrule
    \endhead
    
    \midrule
    \multicolumn{11}{r}{\textit{Continued on next page}} \\
    \midrule
    \endfoot
    
    \bottomrule
    \endlastfoot

    adjnoun & $s$ & 5 & 5 & 5 & 5 & 5 & 5 & 5 & 5 & 2 \\
     & $d$ & 128 & 256 & 128 & 256 & 128 & 128 & 128 & 256 & 128 \\
     & $\rho$ & 0.0001 & 0.0001 & 0.0001 & 0.0001 & 0.0001 & 0.0001 & 0.0001 & 0.0001 & 0.0001 \\\midrule
    bison & $s$ & 2 & 4 & 5 & 4 & 5 & 4 & 4 & 5 & 3 \\
     & $d$ & 256 & 64 & 256 & 128 & 64 & 128 & 64 & 256 & 128 \\
     & $\rho$ & 0.001 & 0.0001 & 0.0001 & 0.0001 & 0.0001 & 0.0001 & 0.0001 & 0.0001 & 0.001 \\\midrule
    blumenau\_drug & $s$ & 5 & 4 & 5 & 5 & 3 & 5 & 5 & 5 & 4 \\
     & $d$ & 64 & 64 & 128 & 256 & 64 & 256 & 64 & 128 & 64 \\
\end{longtable}

\begin{longtable}{lrrrrrrrrrr}
    \caption{\textbf{Optimal LINE$_{1+2}$ Hyperparameters.} For each fraction of removed links and each network we show the optimal hyperparameters used for the results. We created $s \in \left\{2, 3, 4, 5\right\}$ negative samples and set the starting value of the learning rate to $\rho \in \left\{0.0001, 0.001, 0.01\right\}$.
The embeddings were configured with $d \in \left\{64, 128, 256\right\}$ dimensions.} \label{tab:hyperparams_LINE12} \\

    \toprule
    Network & & 0.1 & 0.2 & 0.3 & 0.4 & 0.5 & 0.6 & 0.7 & 0.8 & 0.9 \\
    \midrule
    \endfirsthead
    
    \toprule
    Network & & 0.1 & 0.2 & 0.3 & 0.4 & 0.5 & 0.6 & 0.7 & 0.8 & 0.9 \\
    \midrule
    \endhead
    
    \midrule
    \multicolumn{11}{r}{\textit{Continued on next page}} \\
    \midrule
    \endfoot
    
    \bottomrule
    \endlastfoot

    adjnoun & $s$ & 2 & 5 & 5 & 5 & 2 & 5 & 5 & 5 & 5 \\
     & $d$ & 64 & 128 & 128 & 128 & 256 & 64 & 128 & 256 & 64 \\
     & $\rho$ & 0.001 & 0.0001 & 0.0001 & 0.0001 & 0.001 & 0.0001 & 0.0001 & 0.0001 & 0.01 \\\midrule
    bison & $s$ & 5 & 3 & 5 & 5 & 4 & 5 & 4 & 4 & 2 \\
     & $d$ & 128 & 256 & 128 & 64 & 64 & 256 & 128 & 256 & 256 \\
     & $\rho$ & 0.0001 & 0.0001 & 0.0001 & 0.0001 & 0.0001 & 0.0001 & 0.0001 & 0.0001 & 0.001 \\\midrule
    blumenau\_drug & $s$ & 5 & 5 & 2 & 3 & 5 & 3 & 5 & 3 & 5 \\
     & $d$ & 256 & 128 & 256 & 64 & 256 & 128 & 256 & 128 & 64 \\
\end{longtable}

\begin{longtable}{lrrrrrrrrrr}
    \caption{\textbf{Optimal GCN Hyperparameters.} For each fraction of removed links and each network we show the optimal hyperparameters used for the results. We set the hidden dimensions to $h \in \left\{32, 64, 128\right\}$, the number of layers to $L \in \left\{1, 2, 3\right\}$, the out dimension to $d_{out} \in \left\{64, 128, 256\right\}$ and finally, the learning rate to $\rho \in \left\{0.00001, 0.0001, 0.001, 0.01\right\}$. 
The embeddings were configured with $d \in \left\{64, 128, 256\right\}$ dimensions.} \label{tab:hyperparams_GCN} \\

    \toprule
    Network & & 0.1 & 0.2 & 0.3 & 0.4 & 0.5 & 0.6 & 0.7 & 0.8 & 0.9 \\
    \midrule
    \endfirsthead
    
    \toprule
    Network & & 0.1 & 0.2 & 0.3 & 0.4 & 0.5 & 0.6 & 0.7 & 0.8 & 0.9 \\
    \midrule
    \endhead
    
    \midrule
    \multicolumn{11}{r}{\textit{Continued on next page}} \\
    \midrule
    \endfoot
    
    \bottomrule
    \endlastfoot

    adjnoun & $d$ & 128 & 128 & 256 & 128 & 256 & 256 & 64 & 256 & 256 \\
     & $h$ & 128 & 64 & 64 & 64 & 64 & 64 & 128 & 32 & 128 \\
     & $\rho$ & 0.001 & 0.01 & 0.001 & 0.01 & 0.01 & 0.01 & 0.01 & 0.01 & 0.001 \\
     & $L$ & 2 & 1 & 1 & 1 & 1 & 1 & 1 & 1 & 2 \\
     & $d_{out}$ & 256 & 64 & 128 & 128 & 64 & 128 & 64 & 256 & 128 \\\midrule
    bison & $d$ & 256 & 256 & 128 & 128 & 128 & 64 & 256 & 256 & 128 \\
     & $h$ & 128 & 32 & 64 & 32 & 64 & 32 & 128 & 32 & 64 \\
     & $\rho$ & 0.001 & 0.001 & 0.001 & 0.01 & 0.001 & 0.001 & 0.0001 & 0.01 & 0.01 \\
\end{longtable}

\begin{longtable}{lrrrrrrrrrr}
    \caption{\textbf{Optimal GAT Hyperparameters.} For each fraction of removed links and each network we show the optimal hyperparameters used for the results. We set the hidden dimensions to $h \in \left\{32, 64, 128\right\}$, the number of layers to $L \in \left\{1, 2, 3\right\}$, the out dimension to $d_{out} \in \left\{64, 128, 256\right\}$ and finally, the learning rate to $\rho \in \left\{0.00001, 0.0001, 0.001, 0.01\right\}$. 
The embeddings were configured with $d \in \left\{64, 128, 256\right\}$ dimensions.} \label{tab:hyperparams_GAT} \\

    \toprule
    Network & & 0.1 & 0.2 & 0.3 & 0.4 & 0.5 & 0.6 & 0.7 & 0.8 & 0.9 \\
    \midrule
    \endfirsthead
    
    \toprule
    Network & & 0.1 & 0.2 & 0.3 & 0.4 & 0.5 & 0.6 & 0.7 & 0.8 & 0.9 \\
    \midrule
    \endhead
    
    \midrule
    \multicolumn{11}{r}{\textit{Continued on next page}} \\
    \midrule
    \endfoot
    
    \bottomrule
    \endlastfoot

    adjnoun & $d$ & 128 & 256 & 256 & 128 & 256 & 128 & 256 & 64 & 256 \\
     & $h$ & 32 & 128 & 64 & 32 & 32 & 128 & 32 & 32 & 64 \\
     & $\rho$ & 0.001 & 0.001 & 0.001 & 0.01 & 0.001 & 0.01 & 0.01 & 0.01 & 0.001 \\
     & $L$ & 2 & 1 & 2 & 1 & 1 & 1 & 1 & 1 & 2 \\
     & $d_{out}$ & 128 & 128 & 256 & 256 & 256 & 64 & 128 & 128 & 256 \\\midrule
    bison & $d$ & 256 & 128 & 256 & 128 & 128 & 128 & 128 & 256 & 128 \\
     & $h$ & 32 & 128 & 32 & 128 & 32 & 64 & 32 & 128 & 128 \\
     & $\rho$ & 0.001 & 0.001 & 0.001 & 0.001 & 0.01 & 0.01 & 0.00001 & 0.001 & 0.001 \\
\end{longtable}

\end{document}